\documentclass[aps,prb,twocolumn,showpacs,preprintnumbers,amsmath,amssymb,superscriptaddress]{revtex4}%

\usepackage{graphicx}
\usepackage{dcolumn}
\usepackage{bm}
\usepackage{color}

\begin{document}

\preprint{preprint(\today)}

\title{Cooperative coupling of static magnetism and bulk superconductivity in the
stripe phase of La$_{2-x}$Ba$_{x}$CuO$_{4}$: Pressure- and doping-dependent studies}



\author{Z.~Guguchia}
\email{zurab.guguchia@psi.ch} 
\affiliation{Laboratory for Muon Spin Spectroscopy, Paul Scherrer Institute, CH-5232
Villigen PSI, Switzerland}
\affiliation{Department of Physics, Columbia University, New York, NY 10027, USA}

\author{R.~Khasanov}
\affiliation{Laboratory for Muon Spin Spectroscopy, Paul Scherrer Institute, CH-5232
Villigen PSI, Switzerland}

\author{A.~Shengelaya}
\affiliation{Department of Physics, Tbilisi State University,
Chavchavadze 3, GE-0128 Tbilisi, Georgia}
\affiliation{Andronikashvili Institute of Physics of I.Javakhishvili Tbilisi State University,
Tamarashvili str. 6, 0177 Tbilisi, Georgia}

\author{E.~Pomjakushina}
\affiliation{Laboratory for Developments and Methods, Paul Scherrer Institut, CH-5232 Villigen PSI, Switzerland}

\author{S.J.L.~Billinge}
\affiliation{Condensed Matter Physics and Materials Science Department,
Brookhaven National Laboratory, Upton, NY 11973, USA}

\author{A.~Amato}
\affiliation{Laboratory for Muon Spin Spectroscopy, Paul Scherrer Institute, CH-5232
Villigen PSI, Switzerland}

\author{E.~Morenzoni}
\affiliation{Laboratory for Muon Spin Spectroscopy, Paul Scherrer Institute, CH-5232
Villigen PSI, Switzerland}

\author{H.~Keller}
\affiliation{Physik-Institut der Universit\"{a}t Z\"{u}rich,
Winterthurerstrasse 190, CH-8057 Z\"{u}rich, Switzerland}

\begin{abstract}

 Static spin-stripe order and superconductivity were systematically studied in La$_{2-x}$Ba$_{x}$CuO$_{4}$ (0.11 ${\leq}$ $x$ ${\leq}$ 0.17) at ambient pressure by means of magnetization and ${\mu}$SR experiments. We find that all the investigated La$_{2-x}$Ba$_{x}$CuO$_{4}$ samples exhibit static spin-stripe order and that the quasi two-dimensional superconducting (SC) transition temperature $T_{\rm c1}$ and the static spin-stripe order temperature $T_{\rm so}$ have very similar values throughout the phase diagram. Moreover, the magnetic and the SC properties of the $x$ = 0.155 (LBCO-0.155) and $x$ = 0.17 (LBCO-0.17) samples were studied under hydrostatic pressure. As a remarkable result, in these bulk cuprate superconductors the three-dimensional SC transition temperature $T_{\rm c}$ and $T_{\rm so}$ nearly coincide [$T_{\rm c}(p)$ ${\simeq}$ $T_{\rm so}(p)$] at all pressure investigated (0 ${\leq}$ $p$ ${\leq}$ 2.3 GPa). We also observed a pressure induced transition from long-range spin stripe order to a disordered magnetic state at $p^{\star}$ ${\simeq}$ 1.6 GPa in LBCO-0.155, coexisting with a SC state with substantial superfluid density. In LBCO-0.17 a disordered magnetic state is present at all $p$. The present results indicate that static magnetic order and SC pairing correlations develop in a cooperative fashion in La$_{2-x}$Ba$_{x}$CuO$_{4}$, and provide a new route of understanding the complex interplay between static magnetism and superconductivity in the stripe phase of cuprates.

\end{abstract}


\maketitle

\section{INTRODUCTION}

 Cuprate  high-temperature superconductors (HTSs) have complex phase diagrams with multiple competing ordered phases. 
 One of the most astonishing manifestations of this competition occurs in the system La$_{2-x}$Ba$_{x}$CuO$_{4}$ (LBCO) \cite{Bednorz}, where the bulk superconducting (SC) transition temperature $T_{\rm c}$ exhibits a deep minimum at $x$ = 1/8 \cite{Moodenbaugh,Kivelson,Vojta}. 
At this doping level neutron and x-ray diffraction experiments revealed two-dimensional static charge and spin-stripe order  \cite{Tranquada1,Tranquada2,Abbamonte,HuckerPRB,Luke,Arai}.
While the relevance of stripe correlations for high-temperature superconductivity remains a subject of controversy, the collected  experimental data indicate that the tendency toward uni-directional stripe-like ordering is common to cuprates \cite{Kivelson,Vojta,Kohsaka,Julien}.
Exploring the role of stripe formation for the occurence of high-temperature superconductivity
in cuprates is paramount to elucidate the microscopic pairing mechanism.
  
  On the experimental front, quasi-two-dimensional superconducting correlations were observed in La$_{1.875}$Ba$_{0.125}$CuO$_{4}$ (LBCO-1/8) and La$_{1.48}$Nd$_{0.4}$Sr$_{0.12}$CuO$_{4}$, coexisting with the ordering of static spin-stripes, but with frustrated phase order between the layers \cite{Tranquadareview,Tranquada2008,Li,Valla,Shen}. 
On the theoretical front,  the concept of a sinusoidally modulated [pair-density wave (PDW)] SC order (with the same period as the spin correlations so that its amplitude varies from positive to negative) was introduced, which is intimately intertwined with spatially modulated antiferromagnetism \cite{Berg1,Fradkin,Himeda}.  It has been proposed that both the PDW and the uniform $d$-wave states are close competitors for the SC ground state \cite{Berg1,Fradkin,Himeda}.

 Motivated by the question whether the PDW state is relevant at hole concentrations $x$ $\neq$ 1/8 in La$_{2-x}$Ba$_{x}$CuO$_{4}$, Xu $et.$ $al.$ investigated the system La$_{2-x}$Ba$_{x}$CuO$_{4}$ with $x$ = 0.095 using inelastic neutron scattering \cite{Xu2014}. In this bulk superconductor with $T_{\rm c}$ = 32 K low energy, incommensurate quasi-static antiferromagnetic spin correlations were observed. The coexistence of bulk superconductivity and antiferromagnetic (AFM) spin correlations was explained in terms of a spatially modulated and intertwined pair wave function \cite{Berg1,Fradkin,Himeda,Xu2014}.  There are only a few reports  proposing the relevance of a PDW state in sufficiently underdoped La$_{2-x}$Sr$_{x}$CuO$_{4}$ \cite{Jakobsen2015} and YBa$_{2}$Cu$_{3}$O$_{6-x}$  \cite{Lee,Yu}. At present it is still unclear to what extent PDW order is a common feature of cuprate systems  where stripe order occurs.

   Recently, magnetism and superconductivity in LBCO-1/8, where the stripe order is most stable and magnetism occupies nearly the full volume of the sample, were studied by means of the muon-spin rotation (${\mu}$SR) technique as a function of pressure up to $p$ ${\simeq}$ 2.2 GPa \cite{GuguchiaNJP}.  It was found that application of hydrostatic pressure leads to a remarkable decrease/increase of the magnetic/SC volume fraction. But even at the highest applied pressure the spin order is long-range and occupies a substantial fraction of the sample.  Because of the pressure limit, we were not able to investigate whether it is possible to completely suppress magnetic order and fully restore superconductivity in the stripe phase under pressure. Such an investigation will give important hints for the relevance of the concept of intertwined coexistence of magnetism and superconductivity in striped cuprates.
   
  In this work, static spin-stripe order and superconductivity were systematically studied in polycrystalline  samples of La$_{2-x}$Ba$_{x}$CuO$_{4}$ (0.11 ${\leq}$ $x$ ${\leq}$ 0.17) at ambient pressure by means of magnetization and ${\mu}$SR experiments.  We find that for all the investigated La$_{2-x}$Ba$_{x}$CuO$_{4}$ specimen a substantial fraction of the sample is magnetic and that the 2D SC transition temperature $T_{\rm c1}$ and the static spin-stripe order temperature $T_{\rm so}$ have very similar values throughout the phase diagram. An antagonistic doping dependence of the magnetic volume fraction $V_{m}$ and the diamagnetic susceptibility $\chi_{\rm ZFC}$ was also observed.
Furthermore, we  report on high pressure ${\mu}$SR ($p_{\rm max}$ = 2.2 GPa) and magnetization ($p_{\rm max}$ = 3.1 GPa) studies of the magnetic and superconducting properties of $x$ = 0.155 (LBCO-0.155) and $x$ = 0.17 (LBCO-0.17) samples. We choose these particular compositions ($x$  ${\textgreater}$ 1/8) for the high pressure experiments, since they  are  good superconductors with a well defined single SC transitions and at the same time exhibit magnetic order. LBCO-0.155 also exhibits charge order as previously shown by x-ray and neutron diffraction experiments \cite{HuckerPRB}. 
Remarkably, both $T_{\rm c}$ and $T_{\rm so}$ exhibit a similar pressure dependence in both systems, $i.e.$ $T_{\rm c}(p)$ ${\simeq}$ $T_{\rm so}(p)$, which is an interesting finding. Antagonistic pressure dependence of the magnetic volume fraction $V_{m}$ and the superfluid density ${\rho}_{{\rm s}}$ as well as the diamagnetic moment is also observed, similar to the case of $x$-doping. This suggests phase separation between the SC and the magnetic ground state in La$_{2-x}$Ba$_{x}$CuO$_{4}$. The observed phase separation and the simultaneous appearance of static magnetism and superconductivity in La$_{2-x}$Ba$_{x}$CuO$_{4}$ (0.11 ${\leq}$ $x$ ${\leq}$ 0.17) at ambient pressure and in $x$ = 0.155, 0.17 under hydrostatic pressures indicate that static order and SC pairing correlations develop in a cooperative fashion in La$_{2-x}$Ba$_{x}$CuO$_{4}$, forming a spatially self-organised pattern.

\section{EXPERIMENTAL DETAILS}

\subsection{Sample preparation} Polycrystalline samples of La$_{2-x}$Ba$_{x}$CuO$_{4}$ with $x$ = 0.11, 0.115, 0.125, 0.135, 0.145, 0.15, 0.155, and 0.17 were prepared by 
the conventional solid-state reaction method using La$_{2}$O$_{3}$, BaCO$_{3}$, and CuO. 
The single-phase character of the samples was checked by powder x-ray diffraction.
All the measurements were performed on samples from the same batch. 

\subsection{Instruments}

 ${\mu}$SR experiments under pressure were performed at the GPD instrument (${\mu}$E1 beamline) of the Paul Scherrer Institute (Villigen, Switzerland) \cite{GuguchiaHPI}. The low background GPS (${\pi}$M3 beamline) and Dolly (${\pi}$E1 beamline) instruments were used to study the systems La$_{2-x}$Ba$_{x}$CuO$_{4}$ at ambient pressure.

\subsection{Pressure cells for ${\mu}$SR and magnetization experiments}

The magnetic susceptibility for LBCO-0.155 was measured under pressures up to 3.1 GPa
by a SQUID magnetometer ($Quantum$ $Design$ MPMS-XL). 
Pressures were generated using a diamond anvil cell (DAC) \cite{Giriat} 
filled with Daphne oil which served as a pressure-transmitting medium. 
The pressure at low temperatures was determined 
by detecting the pressure dependence of the SC transition temperature of Pb.
The magnetic susceptibilities for the rest of the samples La$_{2-x}$Ba$_{x}$CuO$_{4}$ were measured only at ambient pressure.\\

 Pressures up to 2.3 GPa were generated in a double wall piston-cylinder type of cell made of MP35N material,
especially designed to perform ${\mu}$SR experiments under pressure
\cite{GuguchiaHPI,Maisuradze,Andreica}. Cells with 6 mm ($p_{max}$ ${\simeq}$ 2.3 GPa) and 7 mm ($p_{max}$  ${\simeq}$ 1.9 GPa) of inner diameters were used for the LBCO-0.155 and LBCO-0.17 samples, respectively. As a pressure transmitting medium Daphne oil was used. The pressure was measured by tracking the SC transition
of a very small indium plate by AC susceptibility. The filling factor
of the pressure cell was maximized. The fraction of the muons stopping
in the sample was approximately 40 ${\%}$ and 50 ${\%}$ for LBCO-0.155 and LBCO-0.17, respectively.
 
\begin{figure*}[t!]
\centering
\includegraphics[width=1.0\linewidth]{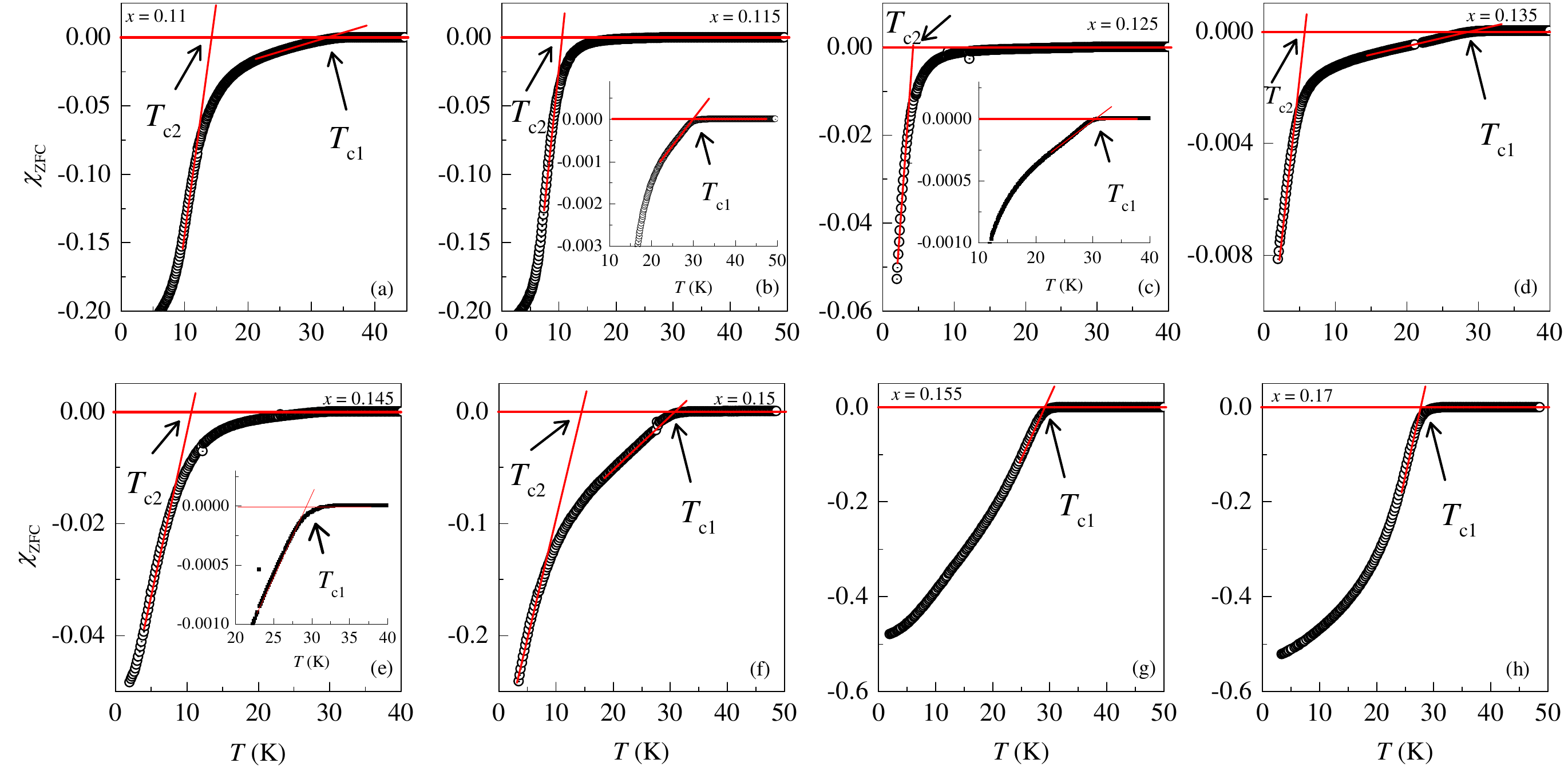}
\vspace{-0.6cm}
\caption{ (Color online) Temperature dependence of the diamagnetic susceptibility $\chi_{\rm ZFC}$ of  La$_{2-x}$Ba$_{x}$CuO$_{4}$ for various $x$, measured at ambient pressures in a magnetic field of $\mu_{0}H$ = 0.5 mT. The arrows denote the superconducting transition temperatures $T_{\rm c1}$ and $T_{\rm c2}$. The insets show the SC transition near $T_{\rm c1}$.} 
\label{fig1}
\end{figure*}

\begin{figure}[t!]
\includegraphics[width=1.0\linewidth]{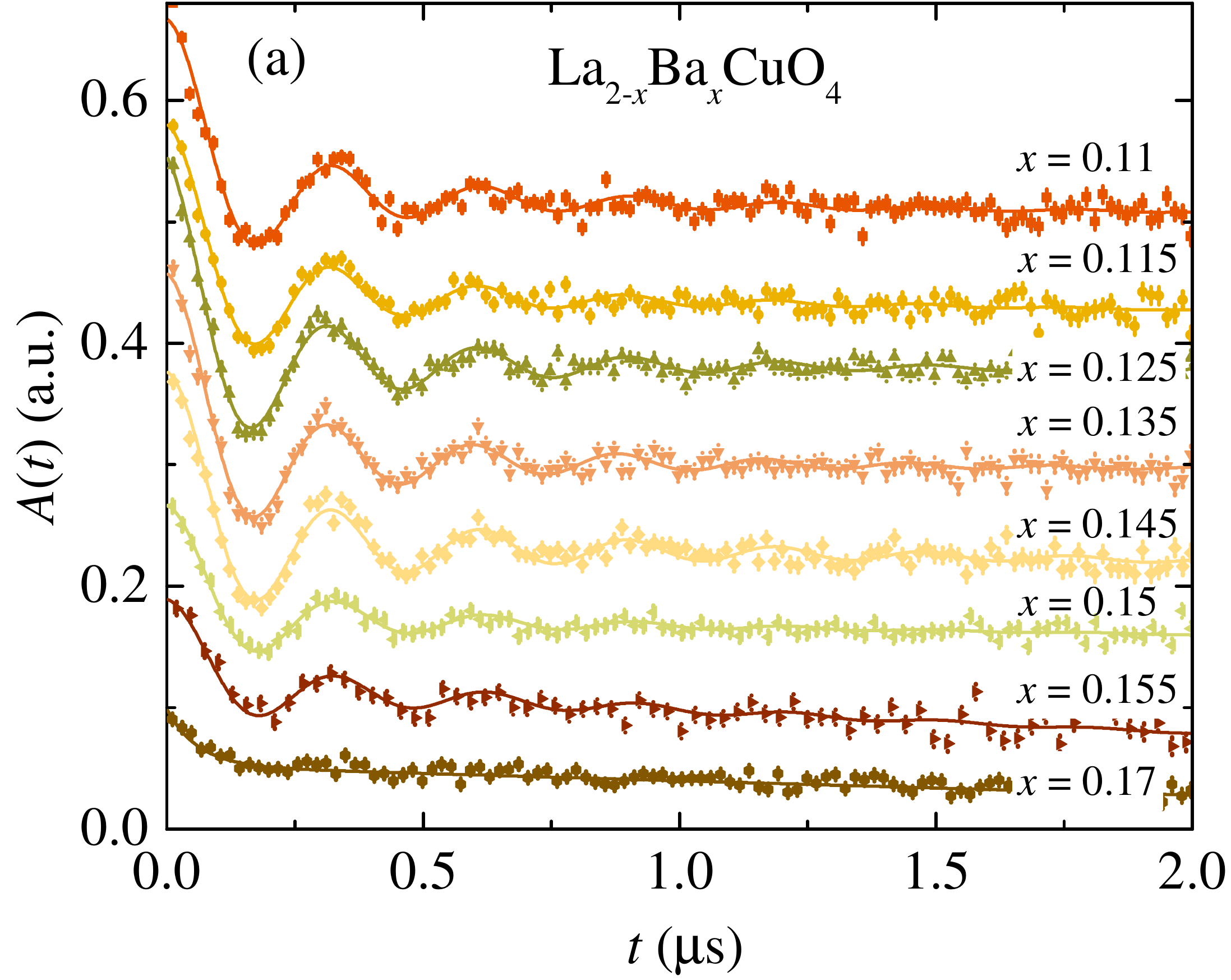}
\includegraphics[width=1.0\linewidth]{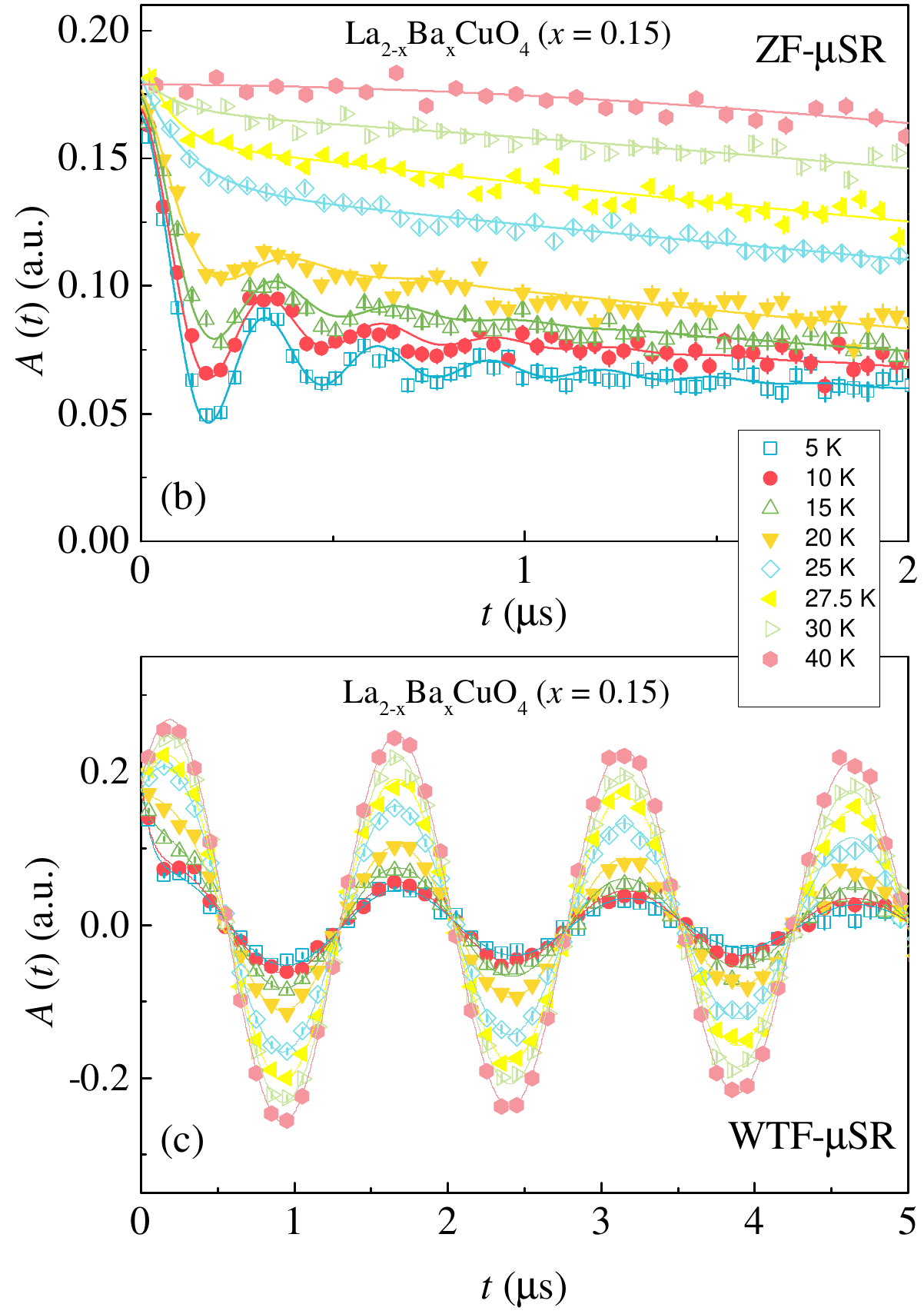}
\vspace{-0.3cm}
\caption{ (Color online) (a) ZF ${\mu}$SR time spectra $A(t)$ for La$_{2-x}$Ba$_{x}$CuO$_{4}$ at various $x$, recorded at 5 K. ZF ${\mu}$SR (b) and WTF ${\mu}$SR (c) spectra for the $x$ = 0.15 sample, recorded at various temperatures.}
\label{fig1}
\end{figure}

\begin{figure}[b!]
\includegraphics[width=1.0\linewidth]{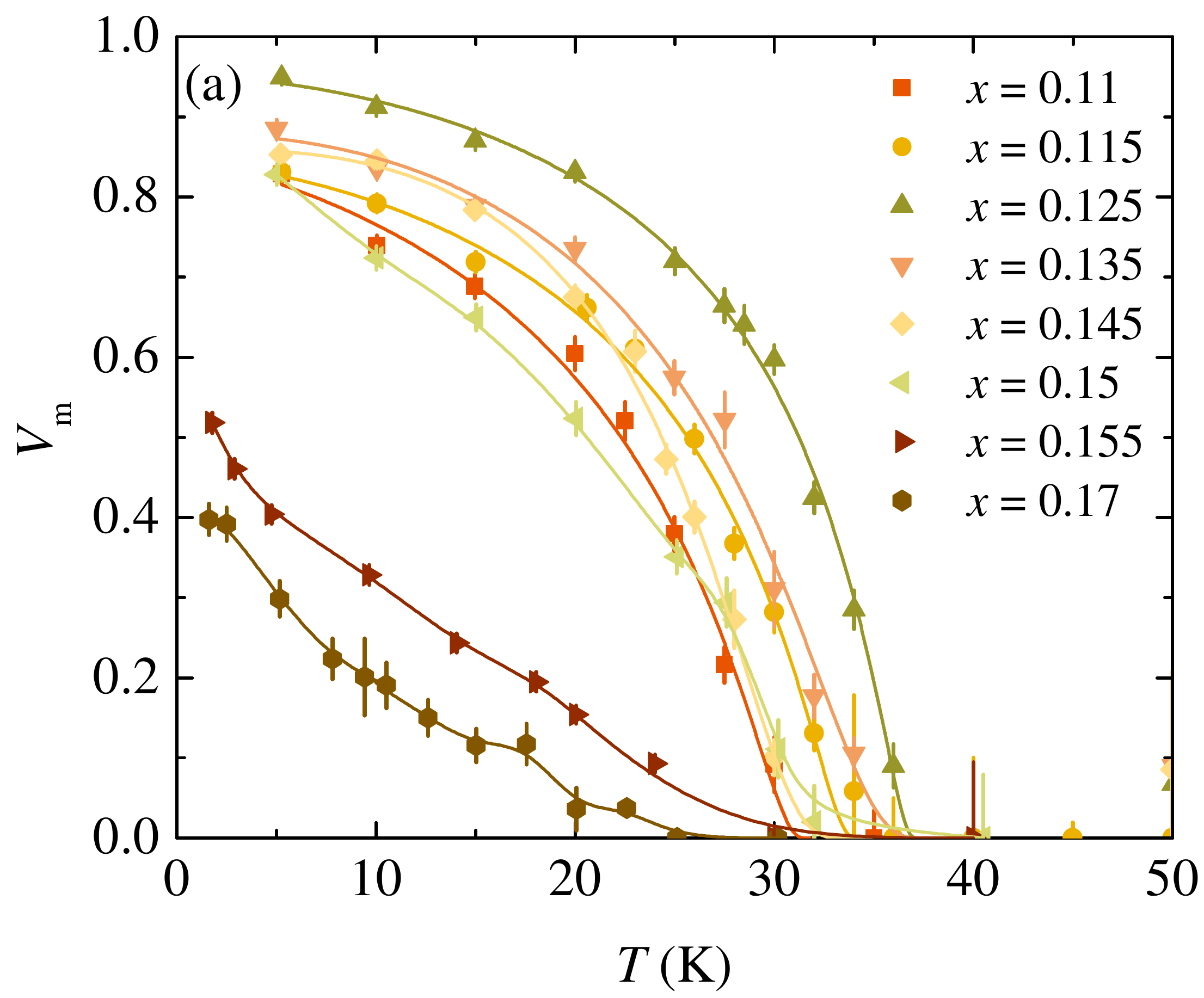}
\includegraphics[width=1.0\linewidth]{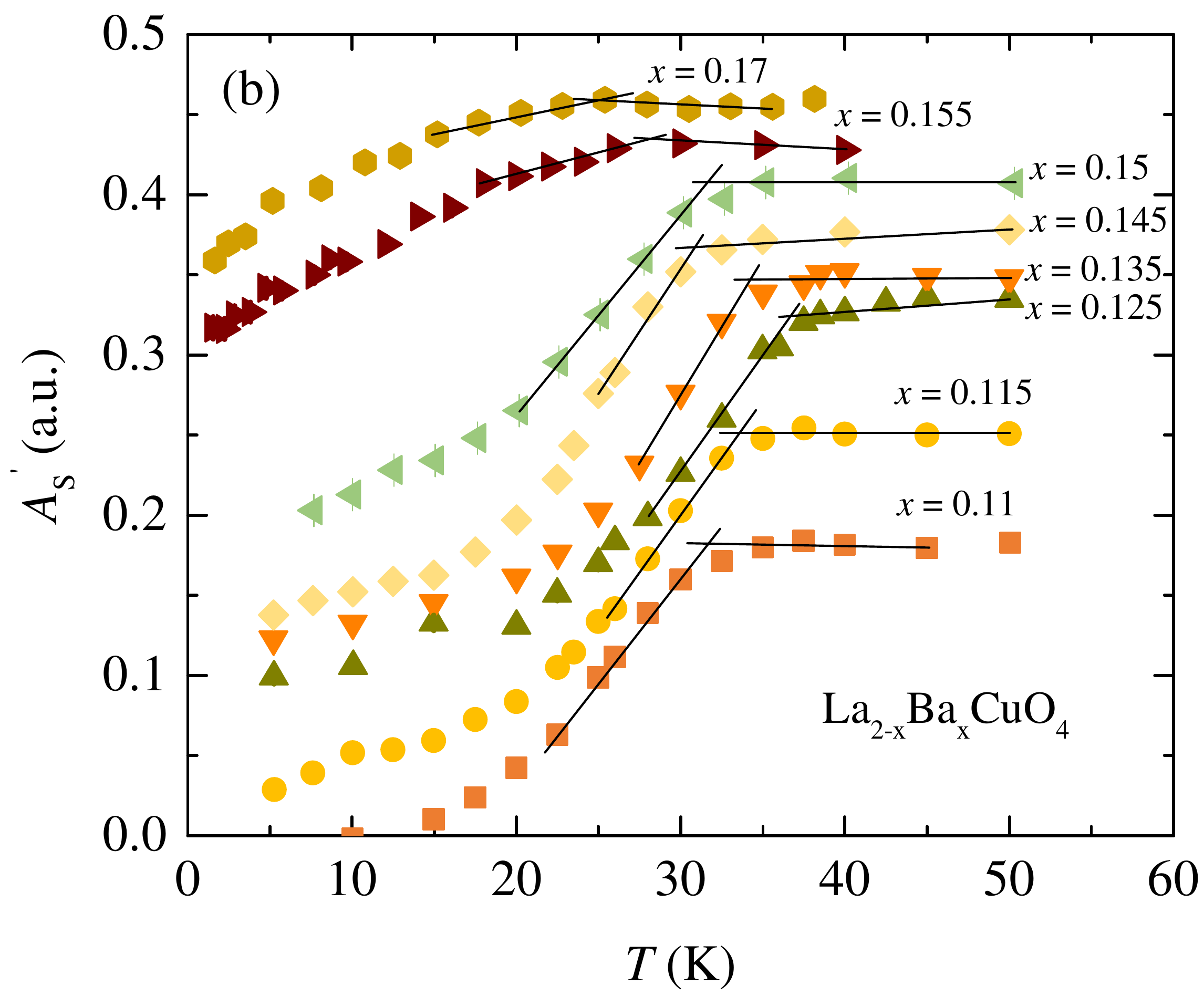}
\vspace{-0.3cm}
\caption{ (Color online) (a) Temperature dependence of the magnetic volume fraction $V_{m}$ in La$_{2-x}$Ba$_{x}$CuO$_{4}$ for various $x$, determined from ZF ${\mu}$SR. (b) The WTF ${\mu}$SR asymmetry for La$_{2-x}$Ba$_{x}$CuO$_{4}$ (0.11 ${\leq}$ $x$ ${\leq}$ 0.17) is plotted as a function of temperature in an applied field of ${\mu}_{0}$$H$ = 3mT. 
The onset temperature to the magnetically ordered state $T_{\rm so}$ is defined as the temperatures where the linearly 
extrapolated low and high temperature data points intersect (indicated by the straight lines).}
\label{fig1}
\end{figure}

\subsection{Analysis of zero-field (ZF) ${\mu}$SR data} 

In the pressure experiment a substantial fraction of the ${\mu}$SR asymmetry originates 
from muons stopping in the MP35N pressure cell surrounding the sample. 
Therefore, the ${\mu}$SR data in the whole temperature range were analyzed by
decomposing the signal into a contribution of the sample and a contribution of the pressure cell:
\begin{equation}
A(t)=A_S(0)P_S(t)+A_{PC}(0)P_{PC}(t),
\end{equation}
where $A_{S}$(0) and $A_{PC}$(0) are the initial asymmetries and $P_{S}$(t) and $P_{PC}$(t) 
are the muon-spin polarizations belonging to the sample and the pressure cell, respectively.
The pressure cell signal was analyzed by a damped Kubo-Toyabe function \cite{Maisuradze}.
The response of the sample consists of a magnetic and a nonmagnetic contribution: 
\begin{equation}
P_S(t)=V_{m}\Bigg[{\frac{2}{3}e^{-\lambda_{T}t}J_0(\gamma_{\mu}B_{\mu}t)}+\frac{1}{3}e^{-\lambda_{L}t}\Bigg]
+(1-V_{m})e^{-\lambda_{nm}t}.
\label{eq1}
\end{equation}
Here, $V_{\rm m}$ denotes the relative volume of the magnetic fraction and $B_{\mu}$ is the average internal magnetic field at the muon site. ${\lambda_T}$ and ${\lambda_L}$
are the depolarization rates representing the transversal and the longitudinal 
relaxing components of the magnetic parts of the sample.
$J_{0}$ is the zeroth-order Bessel function of the first kind.
This is characteristic for an incommensurate spin density wave 
and has been observed in cuprates with static spin stripe order \cite{Nachumi}.
${\lambda_{nm}}$ is the relaxation rate of the nonmagnetic part of the sample.
The total initial asymmetry $A_{\rm tot}$ = $A_{\rm S}$(0) + $A_{\rm PC}$(0) ${\simeq}$ 0.285 is a temperature independent constant. A typical fraction of muons stopped in the sample was $A_{\rm S}$(0)/$A_{\rm tot}$ ${\simeq}$ 0.40(3) and $A_{\rm S}$(0)/$A_{\rm tot}$ ${\simeq}$ 0.50(3), for $x$ = 0.155 and $x$ = 0.17, respectively, which was assumed to be temperature independent in the analysis. The ${\mu}$SR time spectra were analyzed using the free software package MUSRFIT \cite{AndreasSuter}.

\subsection{Analysis of weak transverse field (WTF) ${\mu}$SR data}

The TF-${\mu}$SR spectra were fitted in the time domain with a combination of a slowly
relaxing signal with a precession frequency corresponding to the applied field of $\mu_{0}H$ = 3 mT 
(due to muons in a paramagnetic environment) and a fast relaxing
signal due to muons precessing in much larger static local fields:
\begin{equation}
\begin{split}
A_0P(t)= (A_{PC}e^{-\lambda_{PC} t}  + A_{S}^{'}e^{-\lambda^{'} t}){\cos}(\gamma_{\mu}B^{'}t)+   \\    
A_{S}^{''}\Bigg[{\frac{2}{3}e^{-\lambda_{T}^{''}t}J_0(\gamma_{\mu}B^{''}t)}+\frac{1}{3}e^{-\lambda_{L}^{''}t}\Bigg], 
\label{eq1}
\end{split}
\end{equation}

where $A_{\rm 0}$ is the initial asymmetry, $P(t)$ is the muon spin-polarization function, and $\gamma_{\mu}/(2{\pi}) \simeq 135.5$~MHz/T is the muon gyromagnetic ratio. $A_{PC}$ and ${\lambda}_{PC}$ are the asymmetry and the relaxation rate of the pressure cell signal.
$A_{S}^{'}$ and $A_{S}^{''}$ are the asymmetries of the slowly and fast relaxing sample signals, respectively. 
${\lambda}^{'}$ is the relaxation rate of the paramagnetic part of the sample. ${\lambda}_{T}^{''}$ and ${\lambda}_{L}^{''}$ are the
transverse and the longitudinal relaxation rates, respectively, of the magnetic part of the sample. 
$B^{'}$  and  $B^{''}$ are the magnetic fields, probed by the muons stopped in the paramagnetic and magnetic parts of the sample, respectively.

\begin{figure}[b!]
\includegraphics[width=1.0\linewidth]{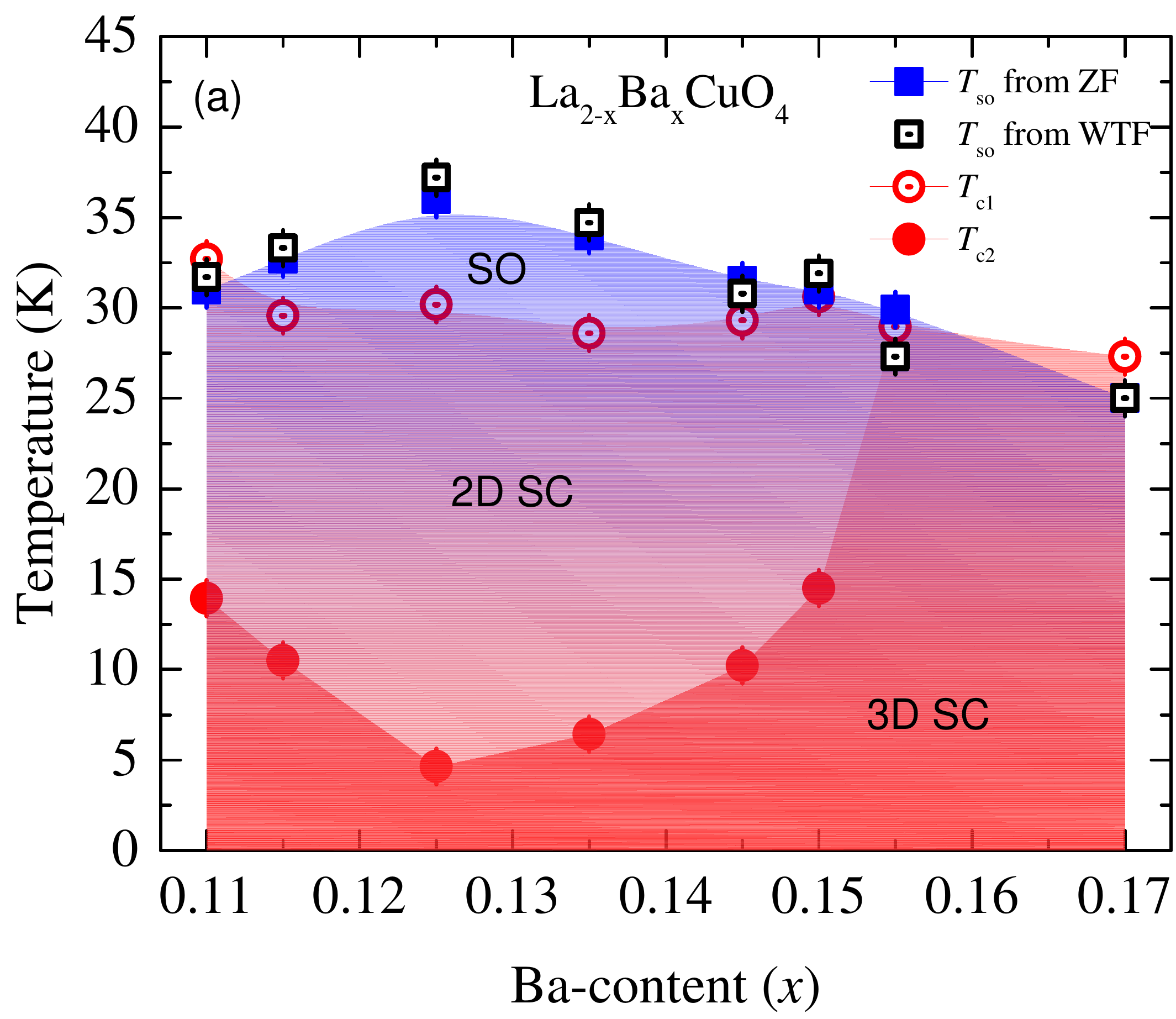}
\includegraphics[width=1.0\linewidth]{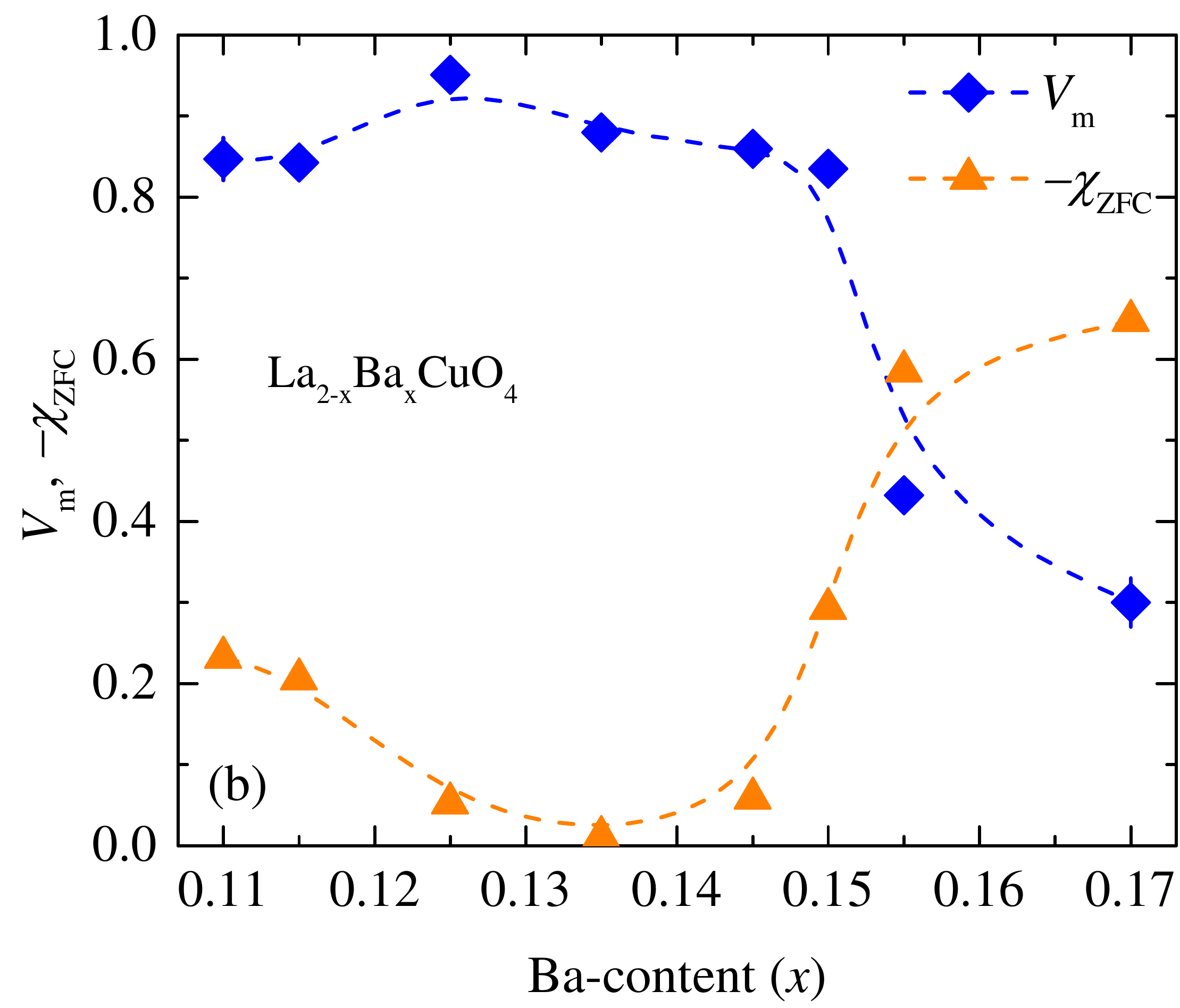}
\vspace{-0.3cm}
\caption{ (Color online) (a) The static spin-stripe order temperature $T_{\rm so}$ and the SC transition temperatures $T_{\rm c1}$ and $T_{\rm c2}$ as a function of Ba-content $x$ in La$_{2-x}$Ba$_{x}$CuO$_{4}$, as determined from ZF-${\mu}$SR, WTF-${\mu}$SR, and magnetization experiments. (b) The magnetic volume fraction $V_{m}$  and the diamagnetic susceptibility $\chi_{ZFC}$ as a function of Ba-content $x$ in La$_{2-x}$Ba$_{x}$CuO$_{4}$.}
\label{fig1}
\end{figure}
 
\begin{figure}[t!]
\centering
\includegraphics[width=1.0\linewidth]{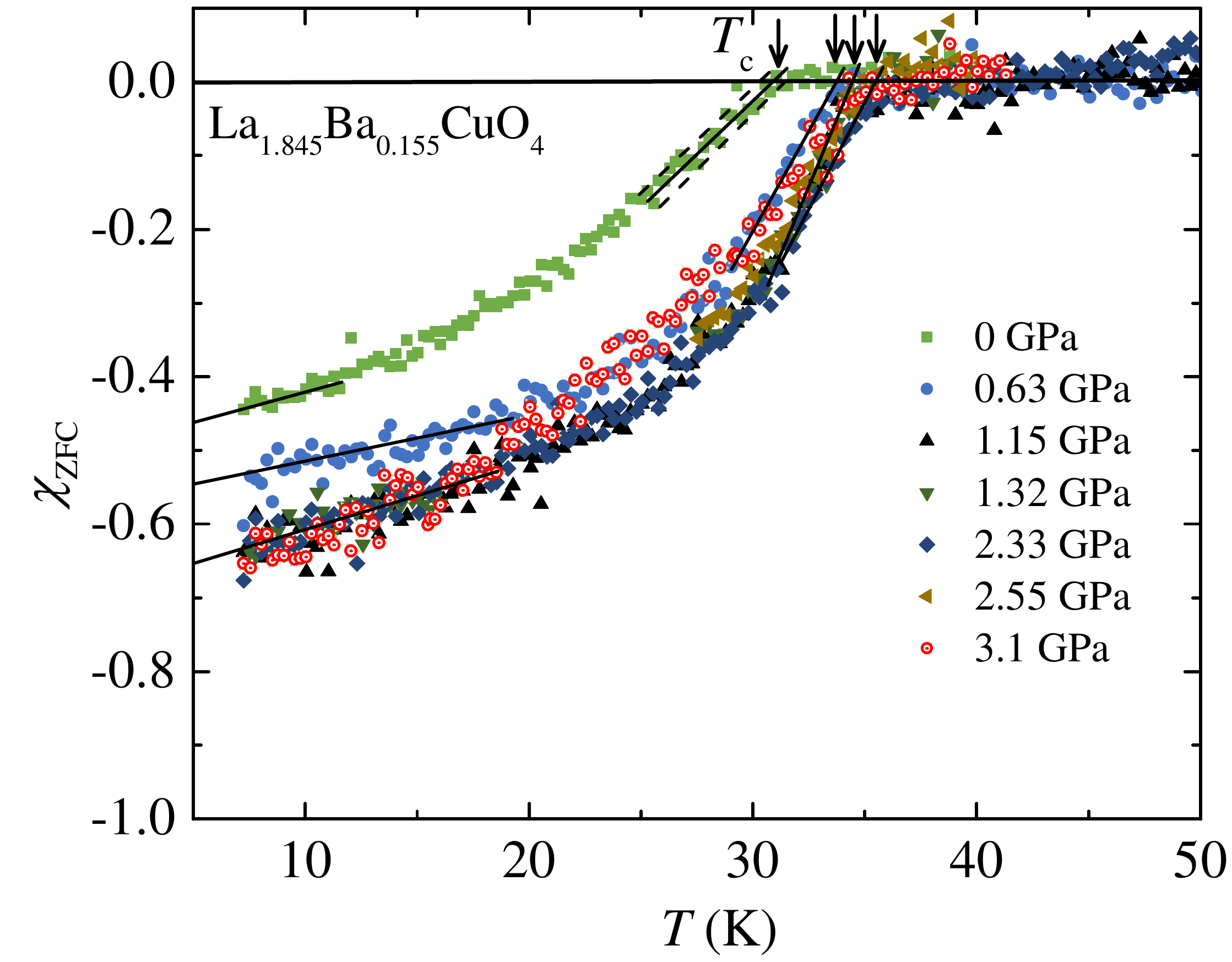}
\vspace{-0.6cm}
\caption{ (Color online) Diamagnetic susceptibility $\chi_{\rm ZFC}$ of LBCO-0.155 as a function
of temperature and pressure. The dependence was measured at ambient and at various applied hydrostatic pressures
in a magnetic field of $\mu_{0}H$ = 0.5 mT. The arrows denote the superconducting transition temperature $T_{\rm c}$.}
\label{fig1}
\end{figure}

\begin{figure}[t!]
\centering
\includegraphics[width=1.0\linewidth]{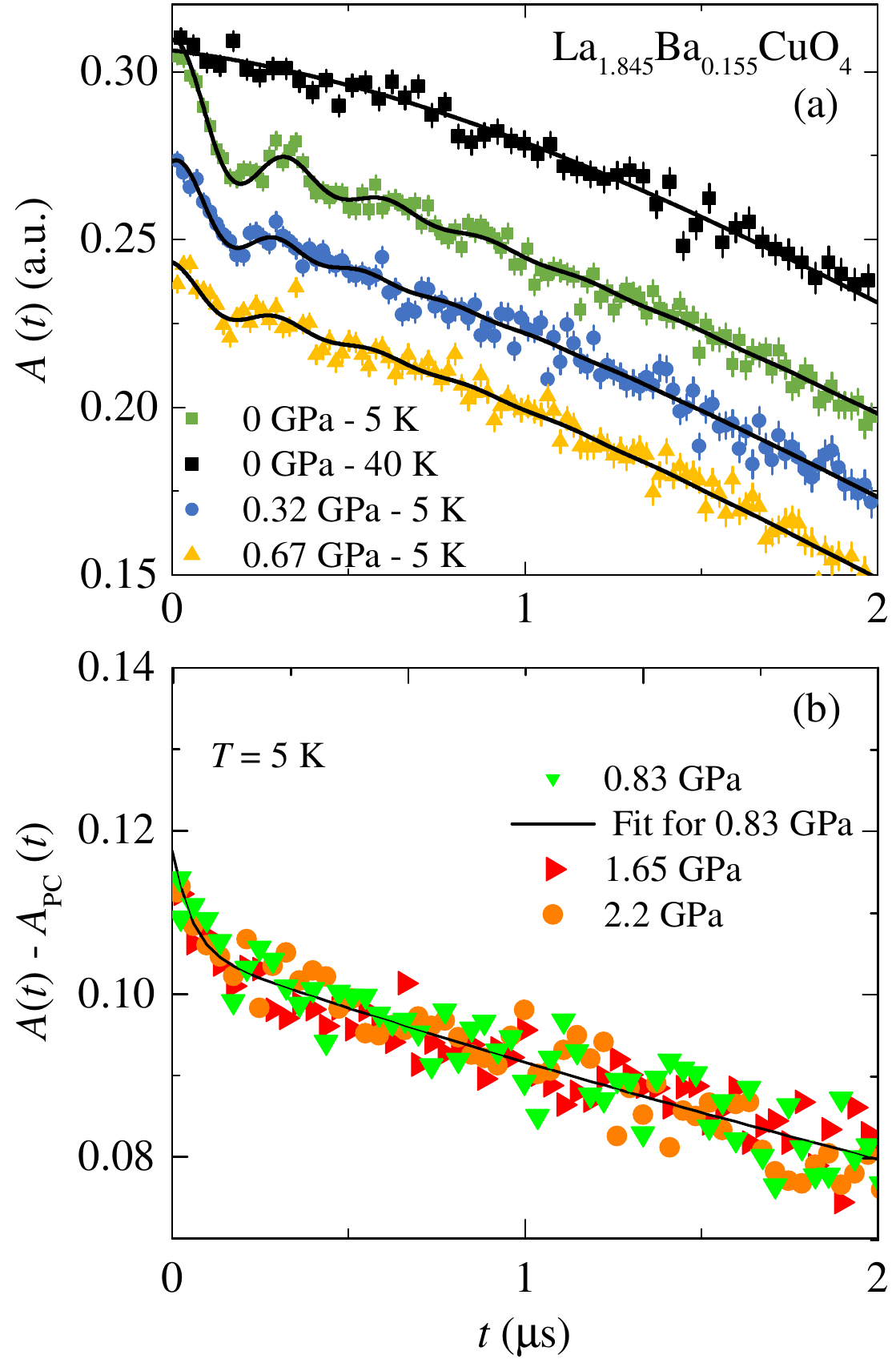}
\vspace{-0.55cm}
\caption{ (Color online) (a) ZF ${\mu}$SR time spectra $A(t)$ for LBCO-0.155 recorded at 5 K in the pressure range of 0 ${\leq}$  $p$ ${\leq}$  0.67 GPa. The ${\mu}$SR time spectrum at 40 K for 0 GPa is also shown. (b)  ${\mu}$SR time spectra for LBCO-0.155 after subtracting the pressure cell contribution from the total signal, 
$A_{S}(t)$ = $A(t)$-$A_{PC}(t)$, recorded at 5 K in the pressure range 0.83 ${\leq}$ $p$ ${\leq}$ 2.2 GPa. The solid lines represent fits to the data by means of Eq.~(1).}
\label{fig1}
\end{figure}

\section{RESULTS}
\subsection{Superconductivity and static spin-stripe order in La$_{2-x}$Ba$_{x}$CuO$_{4}$ (0.11 ${\leq}$ $x$ ${\leq}$
0.17)} 

 Figure 1 shows the temperature dependence of the zero-field-cooled (ZFC) 
magnetic susceptibility $\chi_{\rm ZFC}$ for La$_{2-x}$Ba$_{x}$CuO$_{4}$ (0.11 ${\leq}$ $x$ ${\leq}$ 0.17) samples recorded in a magnetic field of ${\mu}_{\rm 0}$$H$ = 0.5 mT.
For the samples with $x$ = 0.11, 0.115, 0.125, 0.135, 0.145, and 0.15 the diamagnetic moment exhibits a two-step SC transition, as observed in our previous works  for $x$ = 0.125 \cite{GuguchiaNJP,GuguchiaPRL}. For $x$ = 0.125 the first transition appears at $T_{\rm c1}$ ${\simeq}$ 30 K and the second transition at $T_{\rm c2}$ ${\simeq}$ 5 K with a larger diamagnetic response. Detailed investigations performed on single crystalline samples of LBCO-1/8 provided an explanation for this two-step SC transition \cite{Tranquada2008}. The authors interpreted the transition at $T_{\rm c1}$ as due to the development of 2D superconductivity in the CuO$_{2}$ planes, while the interlayer 
Josephson coupling is frustrated by static stripes. A transition to a 3D SC phase
takes place at a much lower temperature $T_{\rm c2}$ ${\ll}$ $T_{\rm c1}$.  
The values of $T_{\rm c1}$ and $T_{\rm c2}$ were defined as the temperatures where the
linearly extrapolated magnetic moments intersect the zero line (see Fig.~1). Note that for the samples with $x$ = 0.155 and 0.17, we observed a well defined single SC transition.

 Figure 2a shows representative zero-field (ZF) ${\mu}$SR time spectra for polycrystalline La$_{2-x}$Ba$_{x}$CuO$_{4}$ (0.11 ${\leq}$ $x$ ${\leq}$ 0.17) samples, recorded at 5 K. The ${\mu}$SR time spectra for all $x$ are well described by a zeroth-order Bessel function which is characteristic for an incommensurate spin density wave,  suggesting the presence of static spin-stripe order in La$_{2-x}$Ba$_{x}$CuO$_{4}$ (0.11 ${\leq}$ $x$ ${\leq}$ 0.17). In a long range ordered magnetic system a coherent muon precession of the whole ensemble is observed giving rise to oscillations in the ZF ${\mu}$SR time spectra as it is the case for all the investigated La$_{2-x}$Ba$_{x}$CuO$_{4}$ samples, except the one with $x$ = 0.17. A damping of the ${\mu}$SR oscillation indicates a distribution of internal magnetic fields sensed by the muon ensemble and is therefore a measure of the disorder in the magnetic system. It is evident that the ${\mu}$SR precession is strongly damped and no coherent precession signal is observed for LBCO-0.17, indicating the presence of a disordered magnetic state in this system. Figure~2b shows the ZF ${\mu}$SR time spectra for the $x$ = 0.15 sample which demonstrates the appearance of the magnetic order below ${\sim}$ 30 K. Figure~3a shows the temperature dependence of the magnetic volume fraction $V_{m}$ extracted from the ZF-${\mu}$SR data for polycrystalline La$_{2-x}$Ba$_{x}$CuO$_{4}$ (0.11 ${\leq}$ $x$ ${\leq}$ 0.17). These data reveal that for all the investigated La$_{2-x}$Ba$_{x}$CuO$_{4}$ specimens a substantial fraction of the sample is magnetic with a relatively high spin-order temperature.
   
 Transverse-field (TF) ${\mu}$SR experiments in weak transverse field (WTF-${\mu}$SR) were also carried out in order to extract $T_{\rm so}$ and compare the values to the ones extracted from the ZF-${\mu}$SR. Figure 2c shows the WTF ${\mu}$SR time spectra for the $x$ = 0.15 sample, which clearly shows the reduction of the amplitude of the ${\mu}$SR signal upon lowering the temperature below ${\sim}$ 30 K, indicating the appearance of the magnetic order. Figure 3b shows the WTF-${\mu}$SR asymmetry for La$_{2-x}$Ba$_{x}$CuO$_{4}$ (0.11 ${\leq}$ $x$ ${\leq}$ 0.17), extracted from the WTF ${\mu}$SR spectra, (following the procedure given in Section II.E) as a function of temperature 
in an applied field of ${\mu}_{0}$$H$ = 3 mT. The onset temperature $T_{\rm so}$ is defined as the temperatures where the linearly extrapolated low and high temperature data points intersect (see Fig. 3).
    
  The values of the static spin-stripe order temperature $T_{\rm so}$ and the SC transition temperatures $T_{\rm c1}$ and $T_{\rm c2}$ for La$_{2-x}$Ba$_{x}$CuO$_{4}$ (0.11 ${\leq}$ $x$ ${\leq}$ 0.17), obtained from susceptibility, ZF ${\mu}$SR, and WTF ${\mu}$SR experiments are summarised in Fig. 4. It is important to note that the values of $T_{\rm so}$ determined from the WTF-${\mu}$SR  experiments are the same as the ones determined from the ZF-${\mu}$SR experiments. This indicates that the WTF-${\mu}$SR measurements give reliable values for $T_{\rm so}$. So this method will be used to extract $T_{\rm so}$ for the samples $x$ = 0.155, 0.17 under pressure.  
$T_{\rm c2}$ shows a local minimum close to 1/8 doping, which is consistent with a previous report \cite{HuckerPRB}. On the other hand, $T_{\rm c1}$ exhibits a high value ${\sim}$ 30 K for all investigated $x$. Note that for $x$ = 0.155 and 0.17
$T_{\rm c1}$ = $T_{\rm c2}$, thus for these two samples the SC transition temperature will be denoted as $T_{\rm c}$ throughout the paper. 
Remarkably, the transition temperatures $T_{\rm c1}$ and $T_{\rm so}$ have very similar values 
throughout the phase diagram, giving strong evidence for a cooperative development of static order and SC pairing correlations in the striped cuprate system La$_{2-x}$Ba$_{x}$CuO$_{4}$. 


 Figure 4b demonstrates an antagonistic doping dependence of the magnetic volume fraction $V_{m}$ and the diamagnetic susceptibility $\chi_{\rm ZFC}$. As it will be shown below for the $x$ = 0.155 sample, $\chi_{\rm ZFC}$(5 K) scales with the magnetic penetration depth $\lambda^{-2}$($T$ = 0) (Shoenberg model). However, a doping induced change of
$\chi_{\rm ZFC}$(5 K)  may be related not only to a change of ${\lambda}$ but also to a change of the SC
volume fraction or combination of both. Since ${\lambda}$ was not measured for whole series of samples La$_{2-x}$Ba$_{x}$CuO$_{4}$, we cannot conclude which one of these two effects plays the dominant role in the observed changes of $\chi_{\rm ZFC}$(5 K). According to the results presented above the samples $x$ = 0.155 and 0.17 exhibit both well defined bulk superconductivity and static spin-stripe order. Furthermore, at ambient pressure static magnetism and superconductivity set in at approximately the same temperature $T_{\rm so}$ ${\simeq}$ $T_{\rm c}$ = 30.5(5) and $T_{\rm so}$ ${\simeq}$ $T_{\rm c}$ = 27.5(5) for LBCO-0.155 and LBCO-0.17, respectively. Therefore we performed investigations of  the stripe order and superconductivity in these systems under pressure. The obtained results will be presented and discussed below.


\subsection{High pressure magnetic susceptibility data}

 Figure 5 shows the temperature dependence of the zero-field-cooled (ZFC) 
magnetic susceptibility $\chi_{\rm ZFC}$ for LBCO-0.155 recorded in a magnetic field of ${\mu}_{\rm 0}$$H$ = 0.5 mT  for selected hydrostatic pressures after substraction  of the background signal from the empty pressure cell.
At ambient pressure superconductivity sets in at $T_{\rm c}$ = 30.5 (5) K (See Fig. 1g).
With increasing pressure $T_{\rm c}$ increases with 3 K/GPa up to $p$ ${\simeq}$ 1.5 GPa where it reaches $T_{\rm c}$ = 35(1) K, then it stays constant up to $p$ = 2.2 GPa. 
For $p$ ${\textgreater}$ 2.2 GPa, $T_{\rm c}$ tends to decrease up to the highest pressure of $p$ = 3.1 GPa.  
The pressure dependence of $T_{\rm c}$ is displayed in Fig. 14 and will be discussed later.
The magnitude of $\chi_{\rm ZFC}$ at the base temperature ($T$ = 5 K) is also enhanced with applied pressure (see Fig.~5) from ${\rvert}{\chi_{\rm ZFC}}{\rvert}$ = 0.45(5) at $p$ = 0 GPa to the saturated value ${\rvert}{\chi_{\rm ZFC}}{\rvert}$ = 0.65(5) at $p$ ${\simeq}$ 1 GPa. It was found that $\chi_{\rm ZFC}$(5 K) scales with $\lambda^{-2}$($T$ = 0) (${\lambda}$ is the magnetic penetration depth) as determined from TF-${\mu}$SR experiments (see Fig. 15a).  According to the Shoenberg model  \cite{Schonberg}, $\chi_{\rm ZFC}$ in a granular sample is expected to scale with $\lambda^{-2}$ due to the penetration of the magnetic field on a distance ${\lambda}$ from the surface of each individual grain. Since we did not measure the grain size for this particular sample it is difficult to judge whether the Schoenberg model applies here and whether the increase of  ${\rvert}\chi_{\rm ZFC}{\rvert}$ is caused by an increase of $\lambda^{-2}$. A pressure induced increase of ${\rvert}\chi_{\rm ZFC}{\rvert}$  can be related either to an increase of the SC volume fraction or to a change of ${\lambda}$ or a combination of both. For all these reasons, it is not possible to extract an absolute value for the SC volume fraction. In LBCO-0.17 superconductivity sets in at $T_{\rm c}$ = 27.5 (5) K  at ambient pressure (See Fig. 1h). Note that for this sample the magnetic susceptibility was not measured under pressure. The SC properties under pressure, {\it i.e.}, $T_{\rm c}$ as well as the superfluid density, were determined by TF ${\mu}$SR experiments (Section III.D).



\subsection{Static magnetism as a function of pressure in LBCO-0.155 and LBCO-0.17}

\begin{figure}[t!]
\centering
\includegraphics[width=1.0\linewidth]{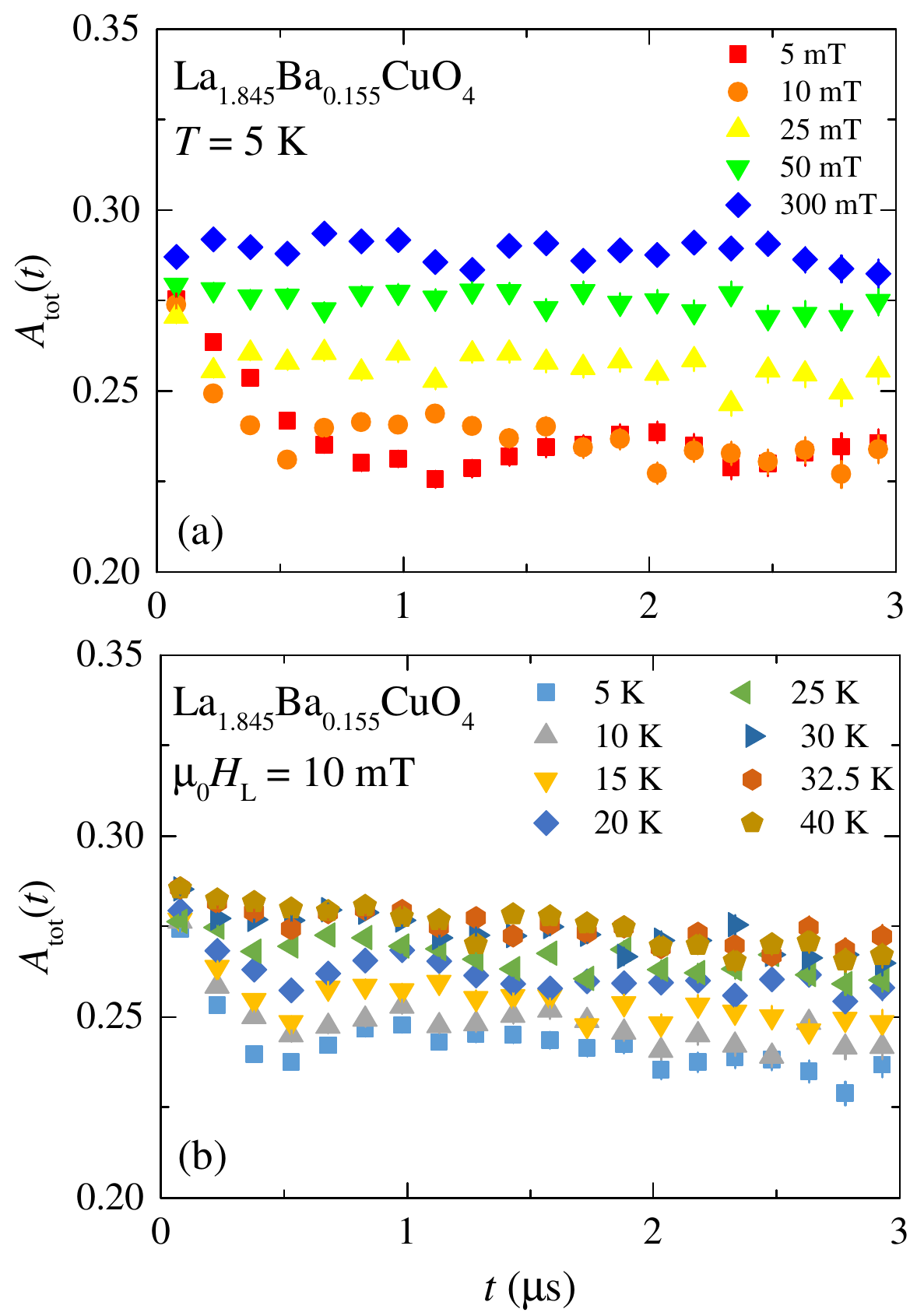}
\vspace{-0.6cm}
\caption{ (Color online) (a) LF-${\mu}$SR time spectra $A_{tot}(t)$ of LBCO-0.155 recorded at $T$ = 5 K and $p$ = 0.83 GPa with different magnetic fields applied along the initial direction of the muon-spin polarization. (b) LF-${\mu}$SR spectra $A_{tot}(t)$ of LBCO-0.155 taken in an applied longitudinal field of $\mu_{0}H_L$ = 10 mT for various temperatures (5 K ${\leq}$ $T$ ${\leq}$ 40 K).}
\label{fig1}
\end{figure}

\begin{figure}[b!]
\centering
\includegraphics[width=1.0\linewidth]{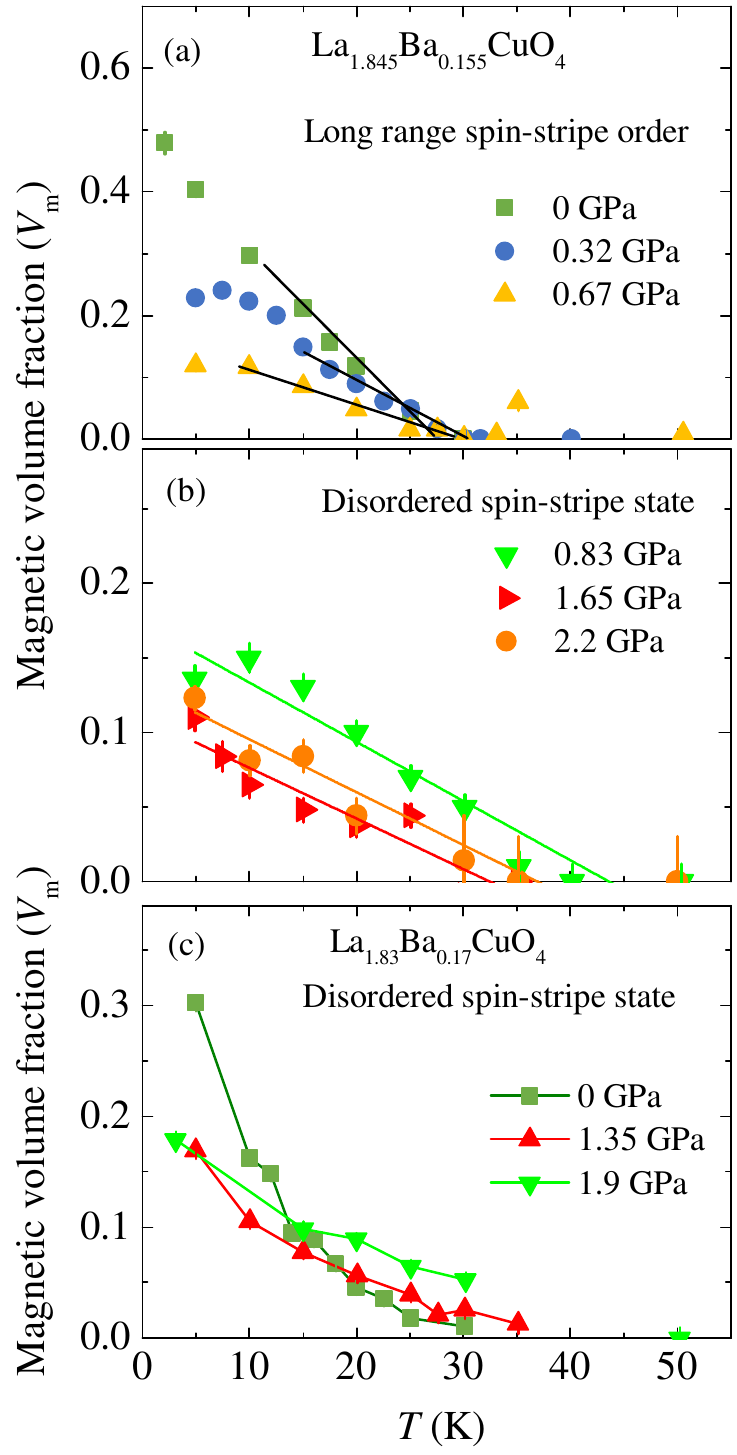}
\vspace{-0.4cm}
\caption{ (Color online) 
Temperature dependence of the magnetic volume fraction $V_{m}$ in LBCO-0.155 
at various hydrostatic pressures: (a) 0 ${\leq}$  $p$ ${\leq}$  0.67 GPa: pressure range where a well defined 
muon spin precession is observed, and (b) 0.83 ${\leq}$  $p$ ${\leq}$  2.2 GPa: pressure range where only a fast depolarization of the ${\mu}$SR time signal is observed. (c) Temperature dependence of  $V_{m}$ in LBCO-0.17 at various hydrostatic pressures. The solid lines are guides to the eye.}
\label{fig1}
\end{figure}

Figures~6a and b show representative ZF ${\mu}$SR time spectra for a polycrystalline LBCO-0.155
sample, recorded at 5 K at ambient  and at selected hydrostatic pressures up to $p$ = 2.2 GPa. 
For ambient pressure the ZF ${\mu}$SR time spectrum taken at 40 K is also shown in Fig.~6a.
At $T$ = 40 K, no muon spin precession, but
only a very weak depolarization of the ${\mu}$SR
signal is observed (see Fig. 6a). This weak depolarization and its
Gaussian functional form are typical for a paramagnetic
material and reflect the occurrence of a small Gaussian Kubo-Toyabe
depolarization, originating from the interaction of the muon spin with randomly oriented nuclear
magnetic moments. At $T$ ${\approx}$ 5 K damped oscillations due to muon-spin precession in internal magnetic fields are observed at pressures up to $p$ = 0.67 GPa. The ${\mu}$SR time spectra are well described by a zeroth-order Bessel function which is characteristic for an incommensurate spin density wave [Eq. ( 2)],  suggesting the presence of long-range static spin-stripe order in LBCO-0.155 \cite{Luke,Nachumi} up to $p$ = 0.67 GPa. The oscillation frequency is nearly pressure independent. On the other hand, for $p$ ${\leq}$ 0.67 GPa the amplitude of this oscillation gradually decreases with increasing pressure, indicating a reduction of the magnetic volume fraction under pressure in agreement with our previous paper \cite{GuguchiaNJP}. In the pressure range 0.83 GPa ${\leq}$  $p$ ${\leq}$ 2.2 GPa, 
instead of the oscillatory behavior seen in the spin-ordered state for $p$ ${\leq}$ 0.67 GPa, a rapidly
depolarizing ZF-${\mu}$SR time spectrum is observed (see Fig. 6b). 
For clarity, Fig. 6b shows  ZF-${\mu}$SR time spectra after subtraction of the pressure cell signal from the total one (Appendix B and Fig. 10). One can clearly see a  rapidly depolarizing component visible at early times (${\leq}$ 0.25 ${\mu}$s) of the spectra, while the non-magnetic part of the sample gives rise to a slow relaxation component,  apparent at longer times (${\textgreater}$ 0.25 ${\mu}$s).
The fast depolarization of the ${\mu}$SR signal (with no trace of an oscillation) could be either due to a broad distribution
of static fields, and/or to strongly fluctuating magnetic moments. To discriminate between these two possibilities,
we have performed decoupling experiments in longitudinal fields. Figure 7a shows $A_{tot}(t)$ measured at 5 K and $p$ = 0.83 GPa with different magnetic fields applied along the initial direction of the muon spin polarization. These experiments show that at modest external fields between 25 and 50 mT (of the order of the internal quasistatic fields) the muon-spin relaxation is substantially suppressed. This means that the muon spins are
fully decoupled from the internal magnetic fields, demonstrating that the weak internal fields are static rather
than dynamic, supporting the quasi-static origin of the fast muon-spin depolarization for $p$ ${\textgreater}$ 0.83 GPa.
The rapid exponential relaxation of the ZF-${\mu}$SR signal implies that the spread of the local magnetic field must be fairly large.  A possible explanation may be that the spatially inhomogeneous magnetic state seen by  ${\mu}$SR  is strongly disordered. This indicates that in LBCO-0.155 the pressure causes a transition from the long-range static spin-ordered to a strongly disordered state at $p$ ${\sim}$ 1 GPa. Figure 7b shows $A_{tot}$(t) recorded in a longitudinal field (LF) of 10 mT at various temperatures. It clearly shows that the muon-spin relaxation increases below about 35 K, providing further evidence for quasi-static magnetic order in LBCO-0.155 at $p$ = 0.83 GPa. For LBCO-0.17 the ${\mu}$SR precession is strongly damped and no coherent precession signal is observed at any pressure, indicating the presence of a disordered magnetic state in this system even at ambient pressure (Fig. 2).

\begin{figure}[t!]
\includegraphics[width=0.9\linewidth]{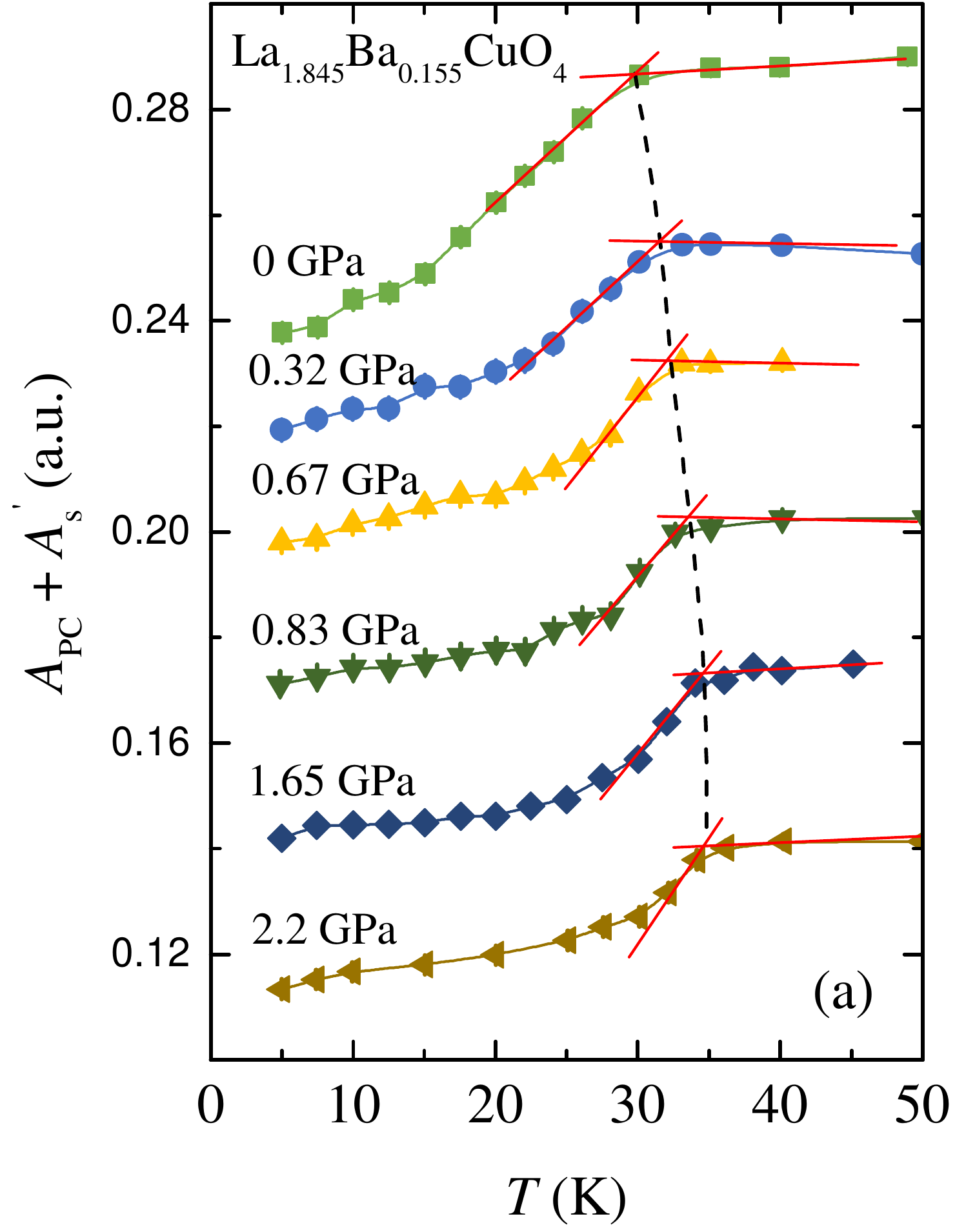}
\includegraphics[width=0.9\linewidth]{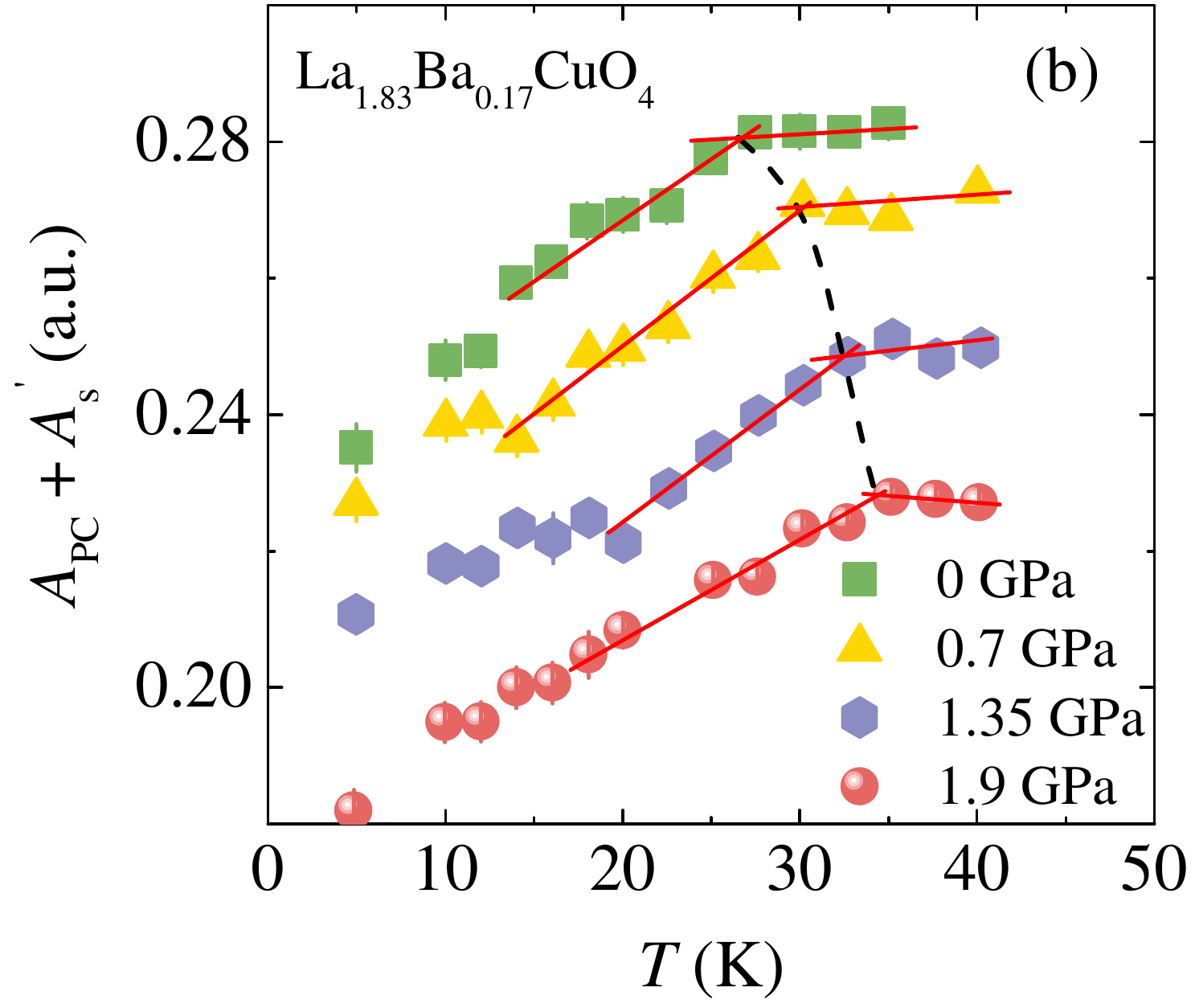}
\vspace{-0.3cm}
\caption{ (Color online) TF asymmetry $A_{S}^{'}$ for LBCO-0.155 (a) and LBCO-0.17 (b) as a function of temperature for ambient and selected applied pressures in an applied field of ${\mu}_{0}$$H$ = 3mT.  The onset temperature to the magnetically ordered state $T_{\rm so}$ is defined as the temperatures where the linearly 
extrapolated low and high temperature data points intersect (indicated by the straight lines). 
The solid curves are guides to the eye.}  
\label{fig1}
\end{figure}

 Next, we present the pressure dependences of the magnetic volume fraction $V_{\rm m}$ and the static spin-stripe order temperature $T_{\rm so}$ in LBCO-0.155 and LBCO-0.17, extracted from the ZF-${\mu}$SR data. Figures ~8a and b show the temperature dependence of $V_{\rm m}$ for LBCO-0.155 in the pressure range of 0 ${\leq}$  $p$ ${\leq}$  0.67 GPa (long-range spin-stripe ordered state) and 0.83 GPa ${\leq}$  $p$ ${\leq}$ 2.2 GPa (disordered spin-stripe state), respectively.
Below $T_{\rm so}$  ${\simeq}$ 30 K, $V_{\rm m}$ increases progressively  with decreasing temperature, and acquires
nearly 50 {\%} at ambient pressure at the base temperature $T$ = 2 K. 
At low temperature $V_{\rm m}$ significantly decreases with increasing pressure, reaching about 15 {\%} at 0.67 GPa (see Fig.~8a). 
On the other hand, $V_{\rm m}$ of the disordered spin-stripe state observed in the pressure range 0.83 GPa ${\leq}$ $p$ ${\leq}$ 2.2 GPa
exhibits a much weaker pressure dependence and reaches about 15 {\%} at $T$ = 5 K (see Fig. 8b). 
Figure 8c shows  the temperature dependence of $V_{\rm m}$ for LBCO-0.17 in the pressure range of 0 ${\leq}$  $p$ ${\leq}$  1.9 GPa (disordered spin-stripe state). Below  $T_{\rm so}$  ${\simeq}$ 25 K, $V_{\rm m}$ increases progressively  with decreasing temperature, and reaches nearly 30 {\%} at ambient pressure at $T$ = 5 K. 

In order to accurately determine $T_{\rm so}$ from ${\mu}$SR, TF-${\mu}$SR experiments in a weak transverse field were carried out. Figures 9a and b show the TF-${\mu}$SR asymmetry $A_S^{'}$ of LBCO-0.155 and LBCO-0.17 extracted from the ${\mu}$SR spectra (following the procedure given in II.E) as a function of temperature 
for ambient and selected applied pressures in an applied field of ${\mu}_{0}$$H$ = 3 mT.
For $p$ = 0 GPa and $T$ ${\textgreater}$  30 K for LBCO-0.155 ($T$ ${\textgreater}$  27 K for LBCO-0.17), $A^{'}$ reaches the maximum value, indicating that the whole sample is in the paramagnetic state, with all the muon spins precessing in the applied magnetic field. Below 30~K (27~K), $A^{'}$ continuously decreases with decreasing temperature.
The reduction of $A^{'}$ signals the appearance of magnetic order in the spin-stripe phase,
where the muon spins experience a local magnetic field larger than the applied magnetic field.
As a result, the fraction of muons in the paramagnetic state decreases.
The onset temperature $T_{\rm so}$ is defined as the temperatures where the linearly
extrapolated low and high temperature data points intersect (see Fig. 9a and b), yielding $T_{\rm so}$ = 30(1) K and $T_{\rm so}$ = 27(1) K 
at $p$ = 0 GPa, for LBCO-0.155 and LBCO-0.17, respectively. Note that this agrees with ZF-${\mu}$SR results.
By applying pressure $T_{\rm so}$ in LBCO-0.155 first increases with increasing pressure, reaching $T_{\rm so}$ = 35(1) K at 1.65 GPa, and 
then tends to saturate as shown in Fig. 14a.  In LBCO-0.17, $T_{\rm so}$ also increases monotonously with increasing pressure, reaching $T_{\rm so}$ ${\simeq}$ 34 K at $p$ = 1.9 GPa (see Fig. 14b). Note that in LBCO-0.15 and LBCO-0.17 $T_{\rm c}(p)$ ${\simeq}$  $T_{\rm so}(p)$ at all applied pressures up to the maximum applied pressure of 2.2 GPa and 1.9 GPa, respectively. This is a remarkable finding.

 \subsection{Probing the vortex state in LBCO-0.155 and LBCO-0.17 as a function of pressure}
%

In the following, we present the pressure dependence of the ${\mu}$SR relaxation rate ${\sigma}_{{\rm sc}}$ for LBCO-0.15 and LBCO-0.17, which is a measure of the superfluid density ${\rho}_{{\rm s}}$ according to the relation: ${\sigma}_{{\rm sc}}$ ${\propto}$  ${\rho}_{{\rm s}}$ ${\equiv}$ $n_{s}/m^{*}$, where $n_{s}$ is the SC carrier density, and $m^{*}$ is the effective mass of the SC carriers.
 
\begin{figure}[t!]
\includegraphics[width=1.0\linewidth]{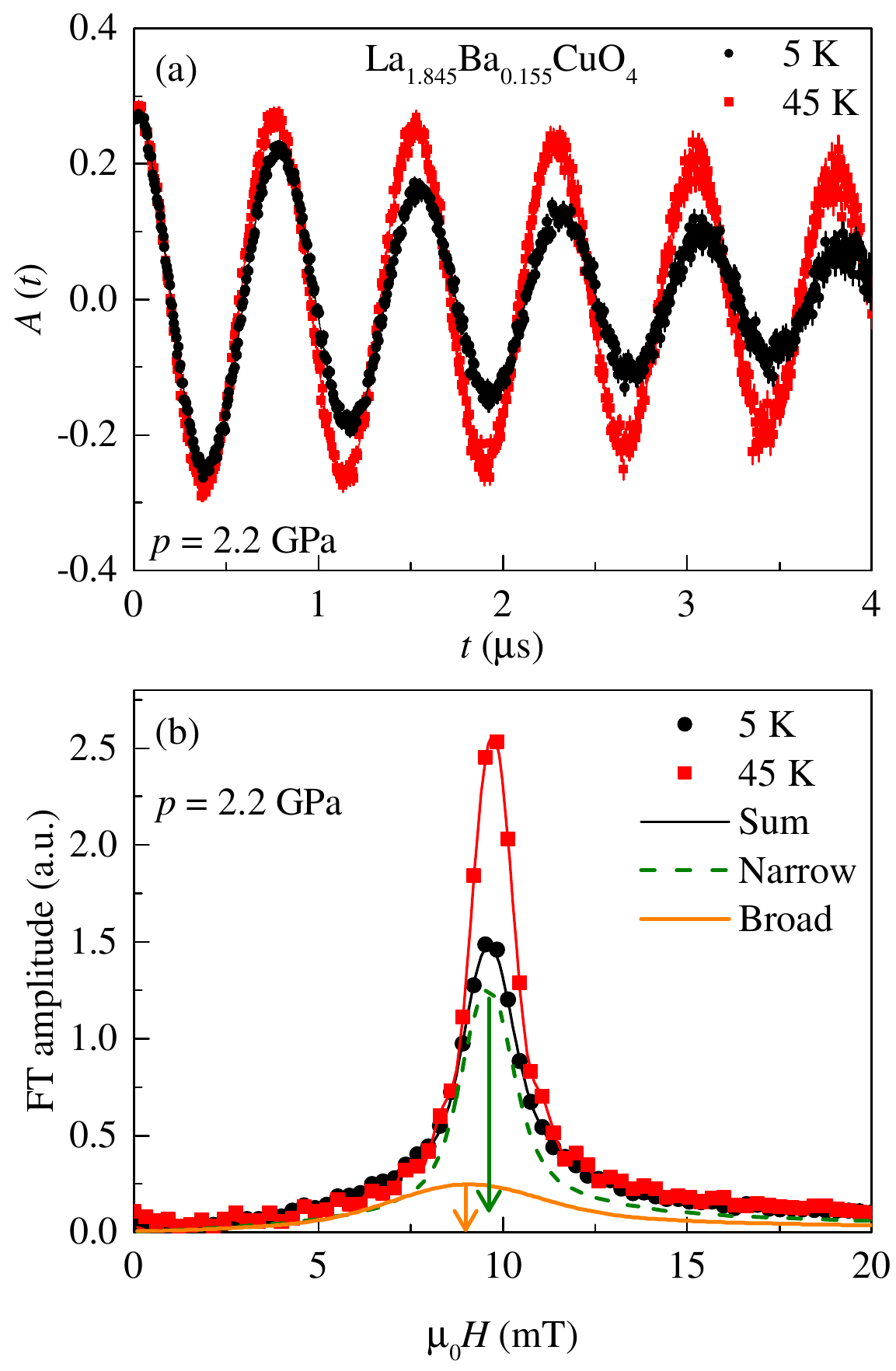}
\vspace{-0.3cm}
\caption{ (Color online) Transverse-field (TF) $\mu$SR time spectra $A(t)$ (a)
and the corresponding Fourier transforms (FTs) (b) of LBCO-0.155.
The spectra were obtained at  $p$ = 2.2 GPa above (45 K) and below (5 K) $T_{{\rm c}}$
(after field cooling the sample from above $T_{{\rm c}}$). The solid lines in panel
(a) represent fits to the data by means of Eq.~(4). The solid lines
in panel (b) are the FTs of the fitted time spectra. The arrows indicate the first moments for the signals of the
pressure cell (green) and the sample (orange), respectively.}  
\label{fig9}
\end{figure}

Figure \ref{fig9}a exhibits TF-${\mu}$SR-time spectra for LBCO-0.155,
measured at the maximum applied pressure $p$
= 2.2 GPa in 10 mT. Spectra above (45 K) and below (5 K) 
the SC transition temperature $T_{{\rm c}}$ are shown. Above $T_{{\rm c}}$
the oscillations show a small relaxation due to the random local fields
from the nuclear magnetic moments. Below $T_{{\rm c}}$ the relaxation
rate strongly increases with decreasing temperature due to the presence
of a nonuniform local magnetic field distribution as a result of the
formation of a flux-line lattice (FLL) in the SC state. Figure \ref{fig9}b
shows the Fourier transforms (FT) of the ${\mu}$SR time spectra
shown in Fig.~\ref{fig9}a. At $T$ = 5 K the narrow signal
around ${\mu}$$_{{\rm 0}}$$H_{\mathrm{ext}}$ = 10 mT  
originates from the pressure cell, while the broad signal with
a first moment $\mu_{0}H_{\mathrm{int}}<\mu_{0}H_{\mathrm{ext}}$,
marked by the orange solid arrow in Fig.~10b, arises from the SC sample.

The TF ${\mu}$SR data were analyzed by using the following functional form: \cite{AndreasSuter} 
\begin{equation}
\begin{aligned}P(t)=A_{s1}\exp\Big[-\frac{(\sigma_{sc}^{2}+\sigma_{nm}^{2})t^{2}}{2}\Big]\cos(\gamma_{\mu}B_{int,s}t+\varphi)\\
+ A_{s2}\Bigg[{\frac{2}{3}e^{-\lambda_{T}t}}+\frac{1}{3}e^{-\lambda_{L}t}\Bigg]  \\
+ A_{pc}\exp\Big[-\frac{\sigma_{pc}^{2}t^{2}}{2}\Big]\cos(\gamma_{\mu}B_{int,pc}t+\varphi),
\end{aligned}
\end{equation}

Here $A_{{\rm s1}}$, $A_{{\rm s2}}$ and $A_{{\rm pc}}$ denote the initial asymmetries of the sample
and the pressure cell, respectively. $A_{{\rm s1}}$ and $A_{{\rm s2}}$ are proportional to the SC and magnetic fractions of the sample, respectively. $\gamma/(2{\pi})\simeq135.5$~MHz/T
is the muon gyromagnetic ratio, ${\varphi}$ is the initial phase
of the muon-spin ensemble. $B_{{\rm int,s}}$ and $B_{{\rm int,pc}}$ represent the local
magnetic fields, probed by the muons, stopped in the sample and the pressure cell, respectively. 
The relaxation rates ${\sigma}_{{\rm sc}}$
and ${\sigma}_{{\rm nm}}$ characterize the damping due to the formation
of the vortex lattice in the SC state and of the nuclear magnetic
dipolar contribution, respectively. In the analysis ${\sigma}_{{\rm nm}}$
was assumed to be constant over the entire temperature range and was
fixed to the value obtained above $T_{{\rm c}}$, where only nuclear
magnetic moments contribute to the ${\mu}$SR relaxation rate ${\sigma}$.
The Gaussian relaxation rate ${\sigma}_{{\rm pc}}$ reflects the depolarization
due to the nuclear magnetism of the pressure cell. It can be seen
from the FTs shown in Fig.~10b that the width of the pressure
cell signal slightly increases below $T_{c}$. As shown previously \cite{Maisuradze-PCm},
this is due to the stray fields in the pressure cell arising from the diamagnetism of the SC sample, 
leading to a temperature dependent ${\sigma}_{{\rm pc}}$
below $T_{c}$. In order to consider this influence, we assume a linear
coupling between ${\sigma}_{{\rm pc}}$ and the field shift of the
internal magnetic field in the SC state: ${\sigma}_{{\rm pc}}$($T$)
= ${\sigma}_{{\rm pc}}$($T$ {\textgreater} $T_{{\rm c}}$) +
$C(T)$(${\mu}_{{\rm 0}}$$H_{{\rm int,NS}}$ - ${\mu}_{{\rm 0}}$$H_{{\rm int,SC}}$),
where ${\sigma}_{{\rm pc}}$($T$ {\textgreater} $T_{{\rm c}}$)
= 0.35 ${\mu}$$s^{-1}$ is the temperature independent Gaussian relaxation
rate. ${\mu}_{{\rm 0}}$$H_{{\rm int,NS}}$ and ${\mu}_{{\rm 0}}$$H_{{\rm int,SC}}$
are the internal magnetic fields measured in the normal and in the
SC state, respectively. As indicated by the solid lines in Figs.~10a and b,
the ${\mu}$SR time spectra are well described by Eq.~(4). The solid lines
in panel (b) are the FTs of the fitted curves shown in Fig.~10a. 
The model used describes the data rather well.

\begin{figure}[t!]
\includegraphics[width=1.0\linewidth]{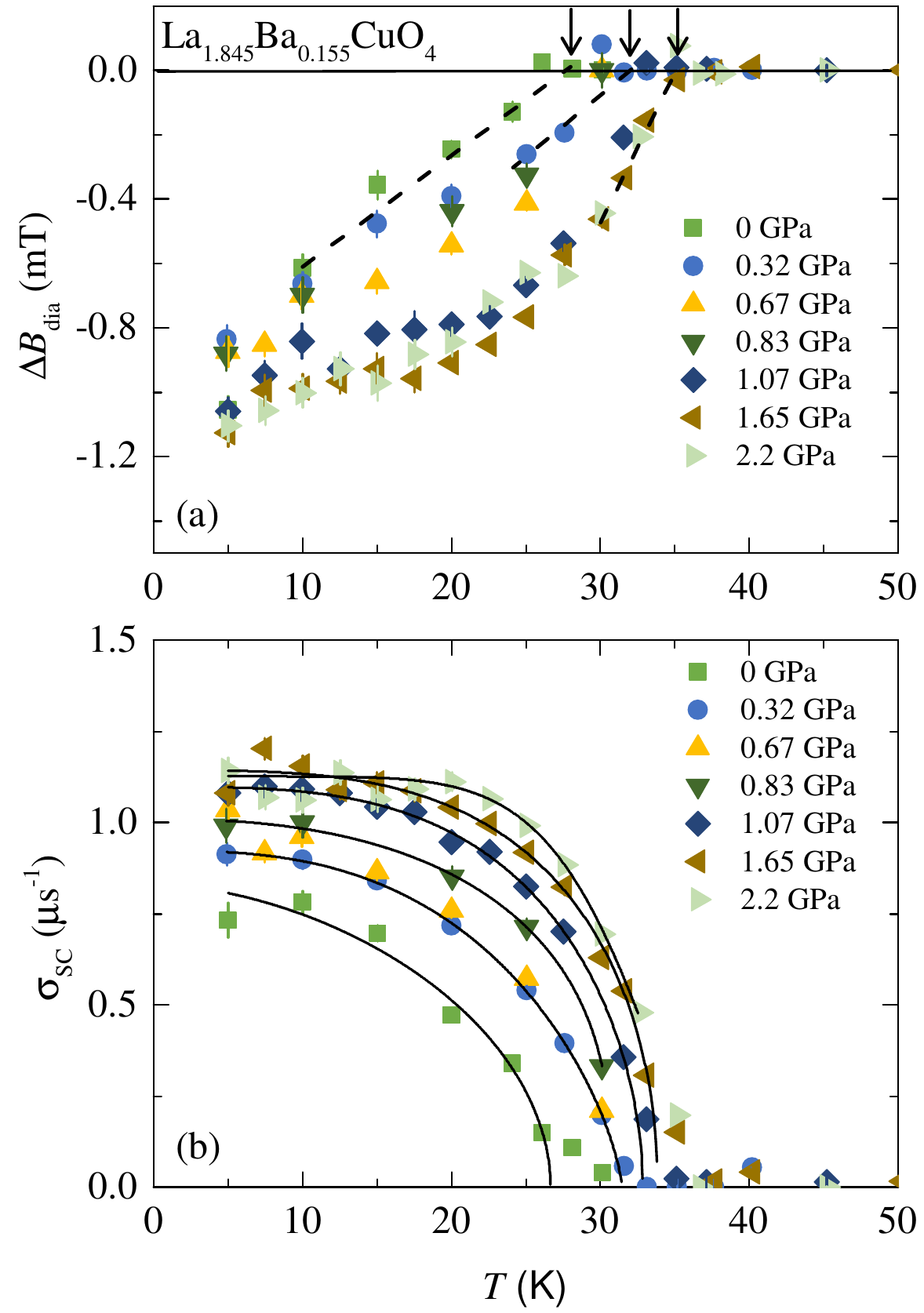}
\vspace{-0.3cm}
\caption{ (Color online) Diamagnetic shift ${\Delta}$$B_{{\rm dia}}$
(a) and ${\mu}$SR relaxation rate ${\sigma}_{{\rm sc}}$ (b) of
LBCO-0.155 as a function of temperature
at various pressures. (a) The definition of the diamagnetic shift
${\Delta}$$B_{{\rm dia}}$ is given in the text. The arrows denote $T_{c}$
for $p$ = 0, 0.32 and 2.2 GPa. (b) ${\mu}$SR
relaxation rate ${\sigma}_{{\rm sc}}$ measured in a magnetic field
of ${\mu_{0}}H$ = 10 mT. The solid lines represent fits of 
the data to the power law described in the text.}  
\label{fig10}
\end{figure}

\begin{figure}[t!]
\includegraphics[width=1.0\linewidth]{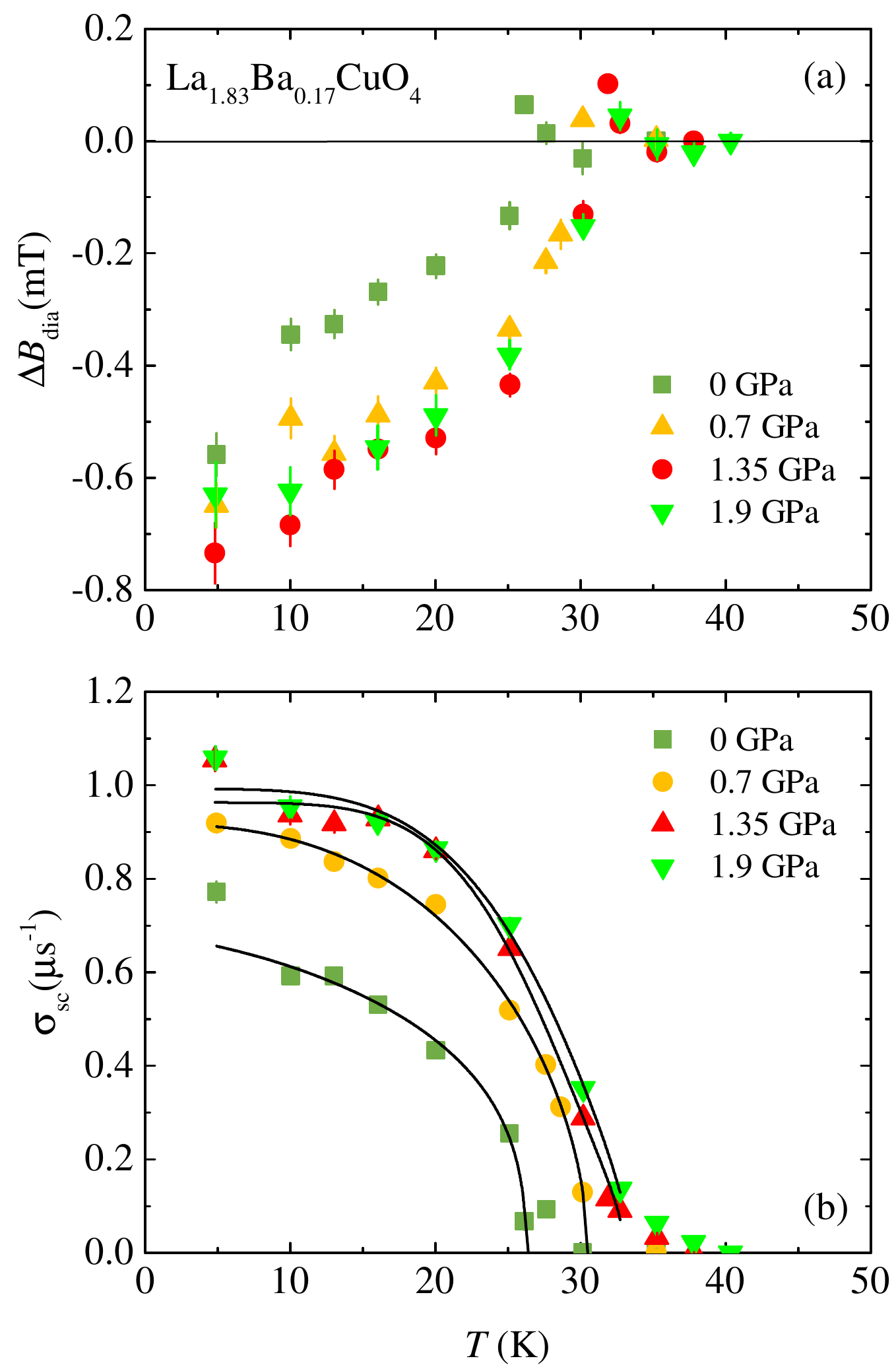}
\vspace{-0.3cm}
\caption{ (Color online) Diamagnetic shift ${\Delta}$$B_{{\rm dia}}$
(a) and ${\mu}$SR relaxation rate ${\sigma}_{{\rm sc}}$ (b) of
LBCO-0.17 as a function of temperature
at various pressures. ${\sigma}_{{\rm sc}}$ is measured in a magnetic field
of ${\mu_{0}}H$ = 10 mT. The solid lines represent fits of 
the data to the power law described in the text.}  
\label{fig10}
\end{figure}

 Below $T_{{\rm c}}$ a large diamagnetic shift of ${\mu}_{{\rm 0}}$$H_{{\rm int}}$
experienced by the muons is observed at all applied pressures. This
is evident in Fig.~11a where we plot the temperature dependence of
the diamagnetic shift ${\Delta}$$B_{{\rm dia}}$ = ${\mu}_{{\rm 0}}${[}$H_{{\rm int,SC}}$-$H_{{\rm int,NS}}${]}
for LBCO-0.155 at various pressures, where ${\mu}_{{\rm 0}}$$H_{{\rm int,SC}}$ denotes the internal field
measured in the SC state and ${\mu}_{{\rm 0}}$$H_{{\rm int,NS}}$
the internal field measured in the normal state at 45 K. Note that ${\mu}_{{\rm 0}}$$H_{{\rm int,NS}}$ is temperature independent.  
The SC transition temperature $T_{{\rm c}}$
is determined from the intercept of the linearly extrapolated ${\Delta}$$B_{{\rm dia}}$
curve and its zero line (we used the same criterium for the determination of $T_{{\rm c}}$ from ${\Delta}$$B_{{\rm dia}}(T)$ as from the susceptibility data ${\chi}_{\rm ZFC}(T)$ presented above), yielding $T_{{\rm c}}$ = 28.3 (5) K for $p$ = 0 GPa.
This value of $T_{{\rm c}}$ is in fair agreement with $T_{{\rm c}}$
= 30.5(9) K obtained from the susceptibility data presented above.
With increasing pressure $T_{{\rm c}}$ increases and reaches $T_{{\rm c}}$ ${\simeq}$ 35 K at $p$ = 1.07 GPa. Above
$p$ = 1.07 GPa, $T_{{\rm c}}$ seems to saturate, which is in 
perfect agreement with the high-pressure magnetization data (see Fig. 5).
Application of pressure causes an enhancement of the diamagnetic shift, which reaches its saturation value above $p$ = 1.07 GPa.

\begin{figure}[t!]
\includegraphics[width=1.0\linewidth]{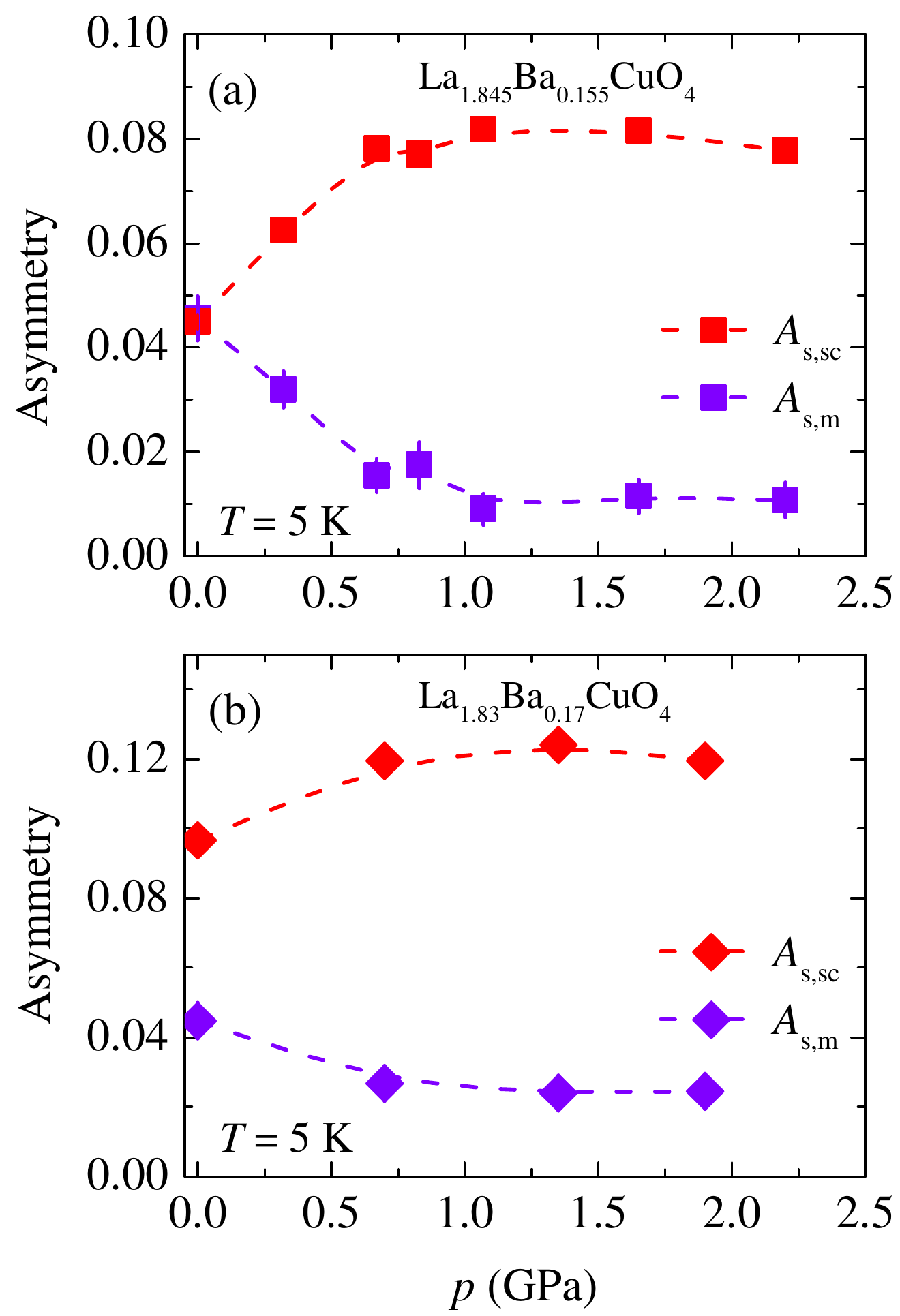}
\vspace{-0.3cm}
\caption{ (Color online) Pressure dependence of the asymmetries $A_{s,sc}$ and $A_{s,m}$ of LBCO-0.17 (a) and LBCO-0.155 (b) at $T$ = 5 K, corresponding to the SC and the magnetic parts of the sample.}
\label{fig11}
\end{figure}


 The temperature dependence of the ${\mu}$SR relaxation rate ${\sigma}_{{\rm sc}}$ 
of LBCO-0.155 in the SC state at selected pressures is shown in Fig.~11b. 
The ${\mu}$SR relaxation rate ${\sigma}_{{\rm sc}}$  is proportional to the
second moment of local magnetic field distribution present in the sample.
Below $T_{{\rm c}}$ the relaxation rate ${\sigma}_{{\rm sc}}$ starts to
increase from zero with decreasing temperature due to the formation
of the FLL. The solid curves in Fig.~11b are fits of the data to the power law 
${\sigma}_{{\rm sc}}$($T$) = ${\sigma}_{{\rm sc}}$(0)[1-($T/T_{\rm c}$)$^{\gamma}$]$^{\delta}$,
where ${\sigma}_{{\rm sc}}$(0) is the zero-temperature value of ${\sigma}_{{\rm sc}}$. 
${\gamma}$ and ${\delta}$ are phenomenological exponents. 
The low-temperature value ${\sigma}_{{\rm sc}}$(0) increases under pressure by about 40 ${\%}$
from $p$ = 0 GPa to $p$ = 1.07 GPa and saturates above (see Fig.~15a). 
Note that the saturation of  the diamagnetic shift ${\Delta}$$B_{{\rm dia}}$ (Fig.~11a), 
diamagnetic susceptibility $\chi_{\rm ZFC}$ (Fig. 15a) as well as of
${\sigma}_{{\rm sc}}$(0) (Fig. 15a) takes place at the same applied pressure $p^{*}$  ${\simeq}$ 1 GPa.

 Figure 12 shows the temperature dependence of the diamagnetic shift ${\Delta}$$B_{{\rm dia}}$ (a) and the ${\mu}$SR relaxation rate ${\sigma}_{{\rm sc}}$ (b) of LBCO-0.177. Application of pressure causes an enhancement of the diamagnetic shift. The low-temperature value ${\sigma}_{{\rm sc}}$(0) increases strongly under pressure up to 0.7 GPa, and then a smooth increase is observed up to the highest appled pressure of $p$ = 1.9 GPa (see Fig. 15b). The overall increase of  
 ${\sigma}_{{\rm sc}}$(0) is about 40 ${\%}$, i.e., same as observed for LBCO-0.155.

 In Figs.~13a and b the asymmetries $A_{s,sc}$ and $A_{s,m}$, which are proportional to the SC and the magnetic fractions of the sample, respectively, are plotted as a function of pressure for LBCO-0.155 and LBCO-0.17. An increase of  $A_{s,sc}$ and a simultaneous decrease of $A_{s,m}$ are observed upon increasing the pressure to $p^{*}$ = 1.07 GPa and 0.7 GPa, for LBCO-0.155 and LBCO-0.17, respectively.  However, above $p^{*}$ both saturates.  This is in excellent agreement with observed pressure dependence of the diamagnetic moment, obtained from the magnetization experiments (see Fig. 15a) and the magnetic volume fraction, extracted from the ZF-${\mu}$SR experiments (see Figs. 15a and b).

\section{DISCUSSION}

\begin{figure}[t!]
\includegraphics[width=1.0\linewidth]{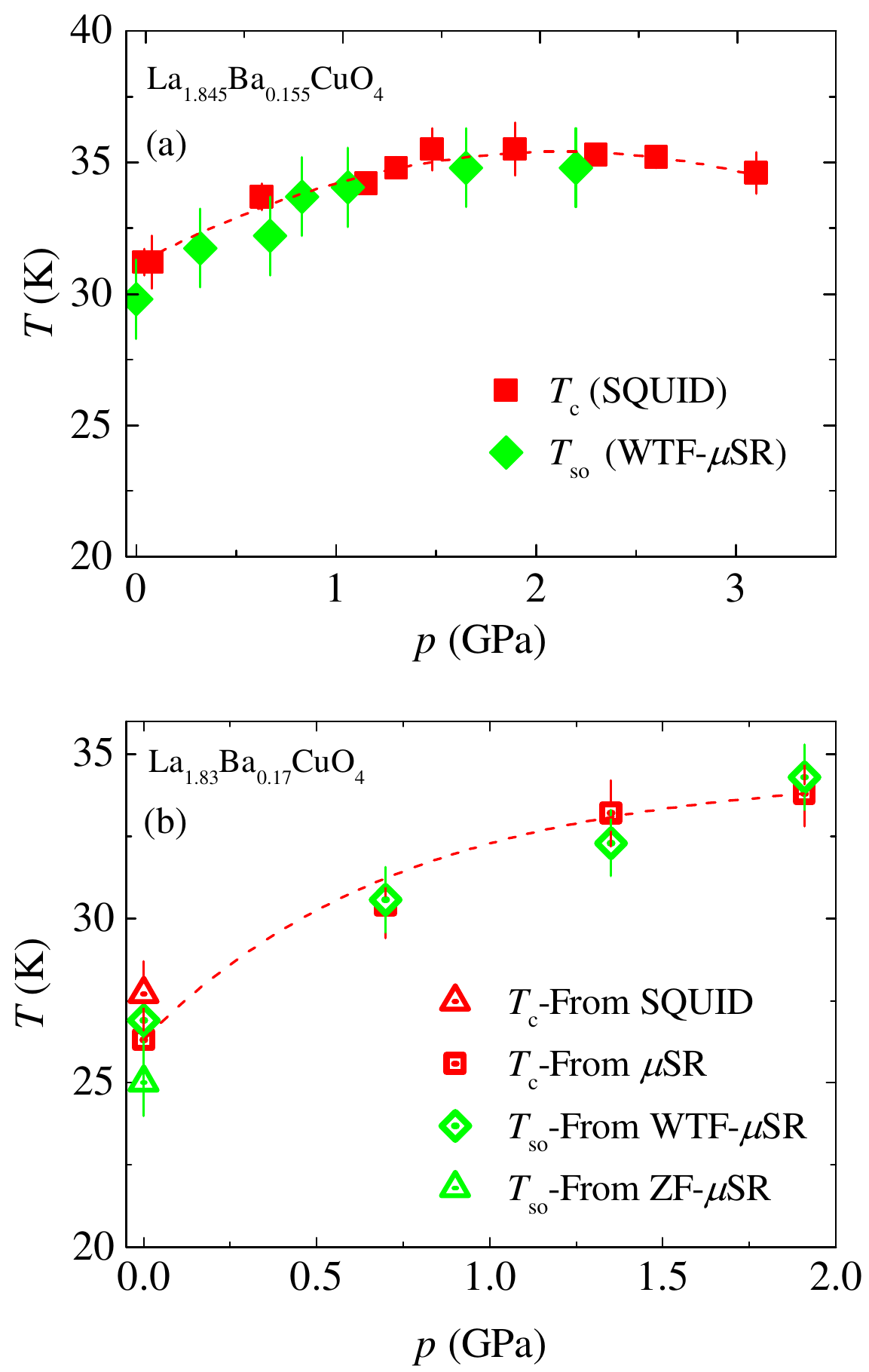}
\vspace{-0.3cm}
\caption{ (Color online) The superconducting transition temperature $T_{\rm c}$ and the magnetic ordering temperature $T_{\rm so}$ of LBCO-0.155 (a) and LBCO-0.17 (b), obtained from DC susceptibility and ${\mu}$SR experiments, are plotted as a function of pressure.}
\label{fig1}
\end{figure}
\begin{figure}[t!]
\includegraphics[width=1.0\linewidth]{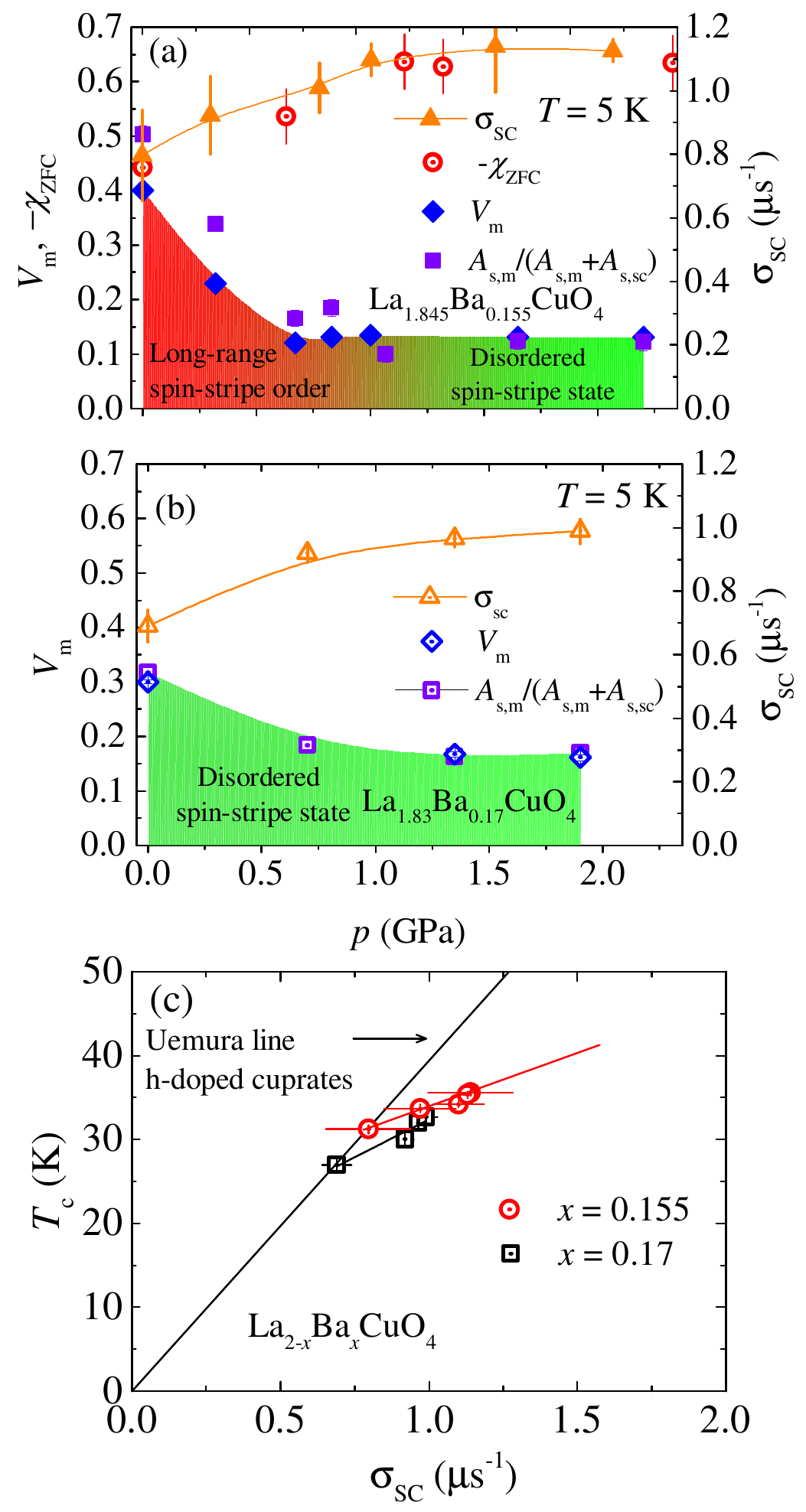}
\vspace{-0.3cm}
\caption{ (Color online) (a) Pressure dependence of the ${\mu}$SR relaxation rate ${\sigma}_{{\rm sc}}$, the diamagnetic susceptibility -$\chi_{\rm ZFC}$, the magnetic volume fraction $V_{m}$ and the quantity $A_{s,m}$/($A_{s,m}$+$A_{s,sc}$) of LBCO-0.155 (a) and LBCO-0.17 (b) taken at 5 K. The solid lines are the  guides to the eye. The long-range spin-stripe order and the disordered spin-stripe state are marked by different colors. (c) Uemura plot ($T_{\rm c}$ vs ${\sigma}_{{\rm sc}}$(0)) for LBCO-0.155 and LBCO-0.17 at zero and applied pressure up to 2.2 GPa and 1.9 GPa, respectively. The Uemura relation observed for underdoped cuprates is represented by the solid line for hole doping. The solid red and black lines represent linear fits of the data.} 
\label{fig1}
\end{figure}

 In order to compare the influence of pressure on the SC and magnetic properties of
LBCO-0.155 and LBCO-0.17, the pressure dependences of the magnetic transition temperature  $T_{\rm so}$, the SC transition temperature $T_{\rm c}$,
and the magnetic volume fraction $V_{\rm m}$ as well as the ${\mu}$SR relaxation rate ${\sigma}_{{\rm sc}}$ 
are plotted in Figs. 14a, 15a and Figs. 14b, 15b for LBCO-0.155 and LBCO-0.17, respectively. In addition, the quantity $A_{s,m}$/($A_{s,m}$+$A_{s,sc}$) is plotted for both samples (Figs. 15a and b).

The most essential findings of the present work are the following: 1) $T_{\rm c}$ and $T_{\rm so}$ have very similar values at all applied pressures as shown in Figs. 14a and b. For LBCO-0.155 both increase up to ${\simeq}$  35(1) K at $p$ ${\simeq}$ 1.5 GPa and are nearly constant at higher pressures up to $p$ ${\simeq}$ 2.2 GPa. 
For LBCO-0.17, $T_{\rm c}$ and $T_{\rm so}$ increase up to ${\simeq}$  34(1) K at the maximum pressure $p$ ${\simeq}$ 1.9 GPa.
2) In LBCO-0.155 pressure causes a transition from the long-range static spin-stripe ordered (0 ${\leq}$ $p$ ${\leq}$ 0.67 GPa) to a strongly disordered ($p$ ${\geq}$ 0.83 GPa) spin-stripe state  (see Fig. 15a). LBCO-0.17 exhibits a disordered magnetic state at ambient as well as at all applied pressures (see Fig. 15b).
To our knowledge this is the first experimental evidence of a similar pressure evolution of  the SC and the magnetic transition temperatures and a pressure induced change of the magnetic state in the stripe phase of cuprates. In view of recent theoretical and experimental works \cite{Mross}, we interprete the observed disordered state as evidence for glassy spin-stripe order. While the absence of long-range stripe order in the presence of disorder is inevitable in two dimensions, a new possible glassy state, the so-called spin-density-wave (SDW) glass state, has recently been proposed \cite{Mross}. In this SDW glass phase the spins are frozen in time, but the phase of the SDW is randomly disordered in space. The spins retain a common axis along which they randomly point up or down (spin nematic order).  In the present case, because pressure does not affect the impurity concentration, the high pressure putative glassy state is most probably caused by a possible frustrated phase separation \cite{EmeryKivelson} between the SC and magnetic ground states in LBCO-0.155, as will be discussed below. \\

\begin{figure}[t!]
\centering
\includegraphics[width=1.0\linewidth]{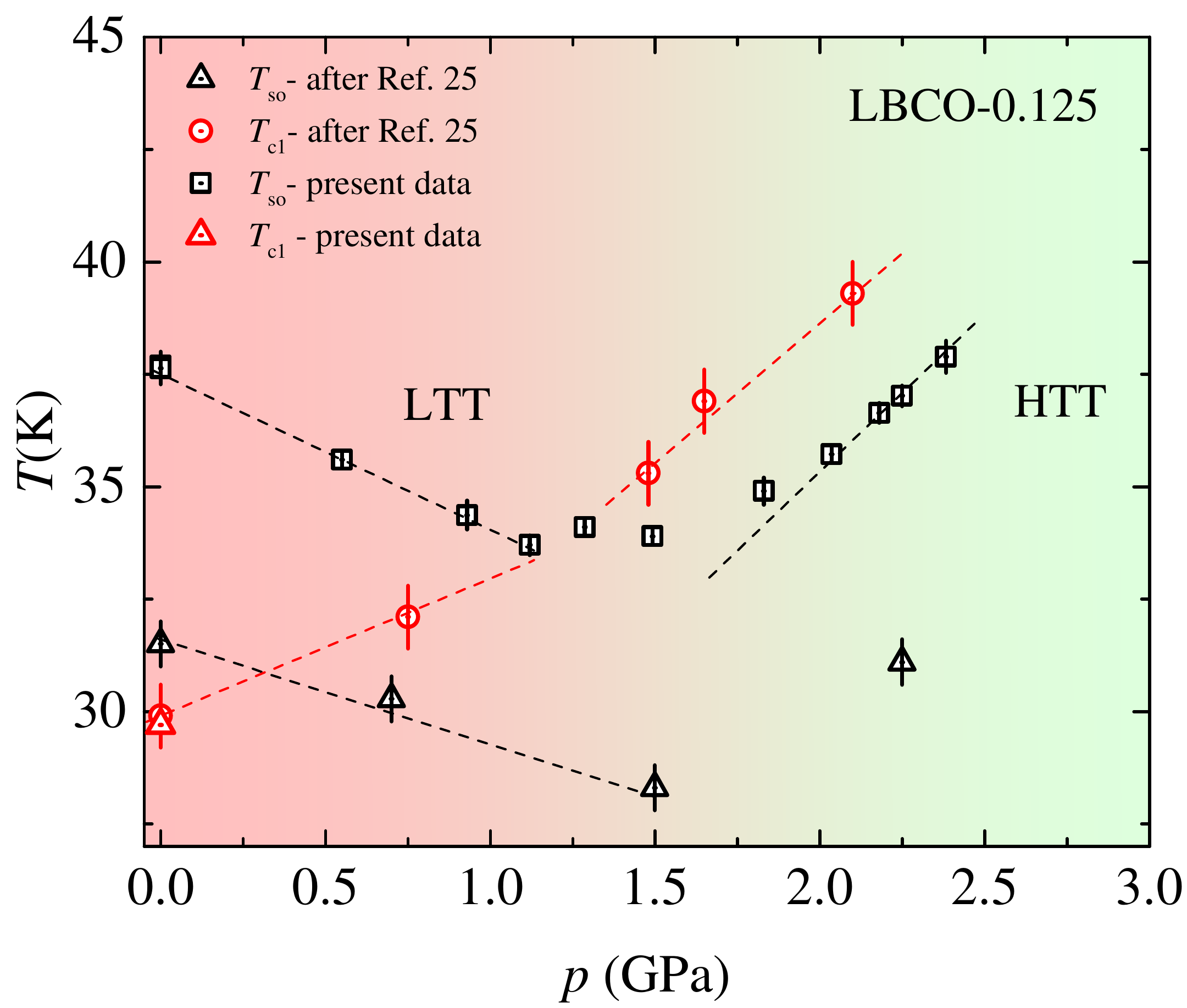}
\vspace{-0.3cm}
\caption{(Color online) Pressure dependence of  $T_{\rm so}$ and $T_{\rm c1}$ for LBCO-1/8, taken from weak-transverse field ${\mu}$SR and magnetization data. The terms LTT and HTT stand for the low-temperature tetragonal and high-temperature tetragonal structural phases.}  
\label{fig10}
\end{figure}

Besides the above results, other interesting findings evident from Figs. 15a and b include the strong increase of the superfluid density ${\rho}_{{\rm s}}$ ${\propto}$ ${\sigma}_{{\rm sc}}$ under pressure and the reduction of the magnetic volume fraction $V_{\rm m}$ under pressure for both LBCO-0.155 and LBCO-0.17, despite the fact that  $T_{\rm so}$ increases with pressure. As shown in Fig. 15a, the low-temperature ($T$ = 5 K) value of ${\sigma}_{{\rm sc}}$ ${\propto}$ ${\rho}_{{\rm s}}$ in LBCO-0.155 increases with increasing pressure  and reaches a constant value at $p$ ${\simeq}$ 1 GPa, which is ${\simeq}$ 35 \% larger than the one at $p$ = 0 GPa. 
On the other hand the magnetic volume fraction $V_{\rm m}$ at $T$ = 5 K sharply decreases with pressure
from ${\simeq}$ 40 \% at ambient pressure to approximately 15 \% at $p$ = 0.83 GPa.
For $p$ ${\textgreater}$ 0.83 GPa $V_{\rm m}$ remains nearly constant. 
As demonstrated in Fig. 15b, LBCO-0.17 shows a similar pressure evolution of ${\sigma}_{{\rm sc}}$ and $V_{\rm m}$, {\it i.e.}, antagonistic pressure behavior between these two quantities. It is interesting to note
that a similar relation was found between the superfluid density and the magnetic volume
fraction in the related compound La$_{1.85-y}$Eu$_{y}$Sr$_{0.15}$CuO$_{4}$ \cite{Kojima}, where 
the tuning of the magnetic and SC properties was realized by rare-earth doping.
To further elucidate the interplay between $V_{\rm m}$ and ${\rho}_{{\rm s}}$, 
it is important to understand the origin for the pressure enhancement of  ${\rho}_{{\rm s}}$.
The Uemura relation in HTSs reflects a remarkable correlation between $T_{\rm c}$ and the zero-temperature 
${\mu}$SR relaxation rate ${\sigma}_{SC}$(0)~${\propto}$~1/${\lambda}^{2}$(0) for cuprate HTSs \cite{Uemura1m}. 
This relation $T_{\rm c}$ vs. ${\sigma}$(0) which  seems to be generic for various families of cuprate HTSs shows in the underdoped regime $T_{\rm c}$ ${\propto}$  ${\sigma}$(0) (Uemura line) (see Fig. 15c).
At higher doping $T_{\rm c}$ saturates and becomes independent of ${\sigma}$(0), and finally in the heavily 
overdoped regime $T_{\rm c}$ is suppressed.  
The initial linear form of the Uemura relation indicates that for these unconventional HTSs the ratio $T_c/E_F$ ($E_{\rm F}$ is an effective Fermi energy) is up to two orders of magnitude larger than for conventional BCS superconductors.  
The Uemura relation for the present ${\mu}$SR pressure data of LBCO-0.155 and LBCO-0.17 are shown in Fig.~15c.
As indicated by the red and the black solid lines the slope $S_{\rm p}$  = (${\delta}$$T_{\rm c}$/${\delta}$$P$)/(${\delta}$${\sigma}_{{\rm sc}}$(0)/${\delta}$$P$) is a factor of ${\simeq}$ 3 and 2 smaller, for LBCO-0.155 and LBCO-0.17, respectively, than that expected from the Uemura line \cite{Uemura1m} with $S_{\rm U}$ = (${\delta}$ $T_{\rm c}$/${\delta}$$P$)/(${\delta}$${\sigma}_{{\rm sc}}$(0)/${\delta}$$P$) ${\simeq}$ 40 K/${\mu}$s. 
A similar substantial deviation from the Uemura line was also observed in previous pressure studies of YBa${_2}$Cu${_3}$O${_7}$ \cite{Maisuradze-PCm} and YBa${_2}$Cu${_4}$O${_8}$ \cite{Khasanov17m,Shengelaya} as well as in oxygen-isotope effect studies of various cuprate superconductors \cite{Khasanov15m}.  A slope which is smaller than that of the Uemura line implies that the increase of the superfluid density ${\rho}_{{\rm s}}$(0) ${\propto}$  ${\sigma}$(0)  is caused not only by an increase of the SC carrier density $n_{s}$, but very likely also by a decrease of the effective mass  $m^{*}$ of the SC carriers. This result together with the observed bulk superconductivity detected by high pressure susceptibility measurements and the similar onset of magnetism and superconductivity in LBCO-0.155 and LBCO-0.17 under pressure suggest that these systems organize themself so as to minimize the overlap between magnetic and superconducting order parameters by intertwining with each other. This suggestion is also supported by the doping dependent studies, revealing the similar values of $T_{\rm so}$ and $T_{\rm c}$ in the series of La$_{2-x}$Ba$_{x}$CuO$_{4}$ samples at ambient pressure (see Fig. 4). This is consistent with the concept of a spatially modulated SC (PDW) state which  may avoid the amplitude-modulated antiferromagnetic spin correlations by intertwining with them \cite{Berg1,Fradkin,Himeda}. Within the scenario of the intertwined orders one may understand that $T_{\rm c}(p)$ ${\simeq}$  $T_{\rm so}(p)$ and the coexistence of the high-pressure magnetic disordered stripe state and bulk superconductivity. Moreover, it also shares the concept of phase separation, but on a short length scale. The frustrated phase separation \cite{Sachdev} between the SC and the long-range magnetic ground states in LBCO-0.155 and LBCO-0.17  leading to a state that is inhomogeneous on an intermediate length scale would be a possible explanation for the antagonistic pressure behaviour between $V_{\rm m}$ and ${\rho}_{{\rm s}}$. This also means that the two coexisting phases are in microscopic proximity to each other.\\

 
Furthermore, one should note the similarities and differences between the pressure effects observed in the present systems LBCO-0.155 and LBCO-0.17 in comparison with LBCO-1/8 \cite{GuguchiaNJP}:
  
 1) At ambient pressure a well defined bulk 3D SC transition with $T_{\rm c}$ ${\simeq}$ 30 K and $T_{\rm c}$ ${\simeq}$ 25 K takes place in  LBCO-0.155 and LBCO-0.17, respectively, while in polycrystalline LBCO-1/8 two SC transitions were observed \cite{GuguchiaNJP,GuguchiaPRL}, as discussed above. 
The first transition appears at $T_{\rm c1}$ ${\simeq}$ 30 K to a quasi 2D SC phase and the second transition at  $T_{\rm c2}$ ${\simeq}$ 5 K corresponds to the transition to a 3D SC phase.

2) $T_{\rm c}$ and $T_{\rm so}$ have very similar values for LBCO-0.155 and LBCO-0.17.
Remarkably,  the values of 2D $T_{\rm c1}$ and $T_{\rm so}$ are also very similar for LBCO-1/8.

3) In the case of LBCO-0.155 and LBCO-0.17 strong positive pressure effects on both $T_{\rm c}$ and $T_{\rm so}$  with $T_{\rm c}(p)$ ${\simeq}$  $T_{\rm so}(p)$ are present. 
We also measured the pressure dependence of the $T_{\rm so}$ in a new LBCO-1/8 sample and we plot the values of $T_{\rm so}$ and $T_{\rm c}$ as a function of pressure in Fig. 16. It is interesting to note that d$T_{\rm so}$/d$p$ and d$T_{\rm c}$/d$p$ have different signs for $p$ ${\leq}$ 1.5 GPa. For $p$ ${\geq}$ 1.5 GPa 
where the so-called low-temperature tetragonal (LTT) phase is suppressed \cite{Hucker,Billinge} both $T_{\rm so}$ and $T_{\rm c}$ increase with increasing pressure with similar slopes. The values of $T_{\rm so}$ and $T_{\rm c1}$ do not really match ($T_{\rm so}$ ${\textless}$ $T_{\rm c1}$), but exhibit the same linear pressure dependence. Note that the values of $T_{\rm so}(p)$ for the new LBCO-1/8 sample are systematically higher that the ones reported previously \cite{GuguchiaNJP}. This can be related to the slightly different preparation procedure of the two samples.


 4) For the samples with $x$ = 1/8 and $x$ = 0.155 we observed an antagonistic pressure dependence of the magnetic volume fraction and the diamagnetic susceptibility. In addition, in the samples with $x$ = 0.155 and 0.17 an antagonistic pressure dependence of  $V_{m}$ and the superfluid density is observed.  
 For the  sample with $x$ = 0.155 we were able to measure both the diamagnetic moment and the superfluid density, and we could compare the changes of both quantities under pressure.
 In the case of $x$ = 1/8 only the diamagnetic moment was measured. We cannot extract reliable information about the superfluid density, since the magnetic fraction is nearly 100 ${\%}$ at the ambient pressure. It is evident that the magnetic fraction decreases with increasing pressure, reaching  50 ${\%}$ at the highest pressure, but it is not possible to follow the pressure evolution of the superfluid density. Probably, one can get a reliable SC response only close to the highest pressure. This means that we cannot definitely conclude which one of the two effects (increase of the SC volume fraction or increase of the penetration depth) plays the dominant role in the enhancement of $\chi_{\rm ZFC}$. Combining all the above mentioned experimental facts we may conclude that magnetism and superconductivity in La$_{2-x}$Ba$_{x}$CuO$_{4}$  are competing phenomena in terms of either volume fraction or superfluid density. In order to discriminate between these two scenarios further experimental and theoretical work is required. The fact that the 2D $T_{\rm c}$ for $x$ =  1/8 as well as the  bulk $T_{\rm c}$ for $x$ = 0.155 and 0.17 have very similar values as $T_{\rm so}$, indicates that the order parameters of superconductivity and magnetism do not compete in terms of pairing strength. This suggests that the cooperative development of static order and SC pairing correlations in the striped cuprate system La$_{2-x}$Ba$_{x}$CuO$_{4}$ may be relevant near 1/8-doping as well as away from it.

 

 The experimental facts listed above point to small differences between the pressure effects of LBCO-1/8 and 
 LBCO-0.155/LBCO-0.17 with $x$ far away from 1/8. To understand these differences between LBCO-1/8  and LBCO-0.155/LBCO-0.17, pressure effects on the structural properties of these systems are crucial. Note that already at ambient pressure the structural properties of these two systems are quite different. While in LBCO-1/8 a discontinuous transition from a low-temperature orthorhombic (LTO) to a LTT phase is observed \cite{HuckerPRB,GuguchiaPRB},  LBCO-0.155 and LBCO-0.17 exhibit no long-range LTT phase\cite{HuckerPRB}. Different structural properties of these two systems may be a possible reason for the differences in the observed pressure effects.

\section{CONCLUSIONS}

  In conclusion, static spin-stripe order and superconductivity were systematically studied in La$_{2-x}$Ba$_{x}$CuO$_{4}$ (0.11 ${\leq}$ $x$ ${\leq}$ 0.17) at ambient pressure by means of magnetization and ${\mu}$SR experiments.  We find that for all investigated doping concerntrations $x$ a substantial fraction of the sample  is magnetic, and the 2D SC transition temperature $T_{\rm c1}$ and the static spin-stripe order temperature $T_{\rm so}$ have very similar values throughout the phase diagram. Moreover, magnetism and superconductivity were studied in LBCO-0.155 and LBCO-0.17 as a function of pressure up to $p$ ${\simeq}$ 3.1 GPa. Remarkably, it was found that in these systems the 3D SC transition temperature $T_{\rm c}$ and $T_{\rm so}$ have very similar values at all pressures, indicating the simultaneous appearance of static magnetic order and superconductivity at all applied pressures in LBCO-0.155 and LBC0-0.17. Antagonistic pressure behaviour between the magnetic volume fraction $V_{m}$ and the superfluid density  ${\rho}_{{\rm s}}$ was observed in LBCO-0.155 and LBC0-0.17 under pressure up to $p$ ${\simeq}$ 0.83 GPa and $p$ ${\simeq}$ 0.7 GPa, respectively, which was interpreted in terms of a frustrated phase separation scenario. Interestingly, in LBCO-0.155 for $p$ ${\geq}$ 0.83 GPa, where ${\rho}_{{\rm s}}$ reaches a constant value, long-range static spin-stripe order is not suppressed, but is replaced by a disordered quasi-static magnetic state, which persists up to the highest applied pressure of $p$ = 2.2 GPa. A disordered static magnetic state, observed in LBCO-0.17 at ambient pressure, also persists up to the highest applied pressure of $p$ = 1.9 GPa. The present findings provide clear experimental evidence of a pressure induced change of the magnetic state as well as the same pressure evolution of  the SC and the magnetic transition temperatures in the stripe phase of cuprates. These experimental results strongly suggest that static spin-stripe order and SC pairing correlations develop in a cooperative fashion in La$_{2-x}$Ba$_{x}$CuO$_{4}$.



 \textbf{\section{Acknowledgments}}

The ${\mu}$SR experiments were performed at the Swiss Muon Source (S${\mu}$S)
Paul Scherrer Insitute, Villigen, Switzerland. Z.G. thanks Y.J. Uemura, S.A. Kivelson and J. Tranquada for helpful discussions. Z.G. thanks P.K. Biswas for his technical support during the experiments on Dolly ${\mu}$SR Instrument. Z.G. gratefully acknowledges the financial support by the Swiss National Science Foundation (SNFfellowship P2ZHP2${\_}$161980 and SNFGrant 200021${\_}$149486). A.S. acknowledges support from the SCOPES grant No. Z74Z0${\_}$160484. We further thank A. Schilling and F.v. Rohr for supporting the susceptibility measurements of LBCO-0.155 under pressure. Work in the Billinge group was supported by U.S. Department of Energy, Office of Science, Office of Basic Energy Sciences (DOE-BES) under contract No. DE-SC00112704.

\newpage 

\subsection{Supplementary Information: Extracting the ZF-${\mu}$SR signal  of LBCO-0.155 for $p$ ${\geq}$ 0.83 GPa}
 
\begin{figure}[b!]
\centering
\includegraphics[width=1.0\linewidth]{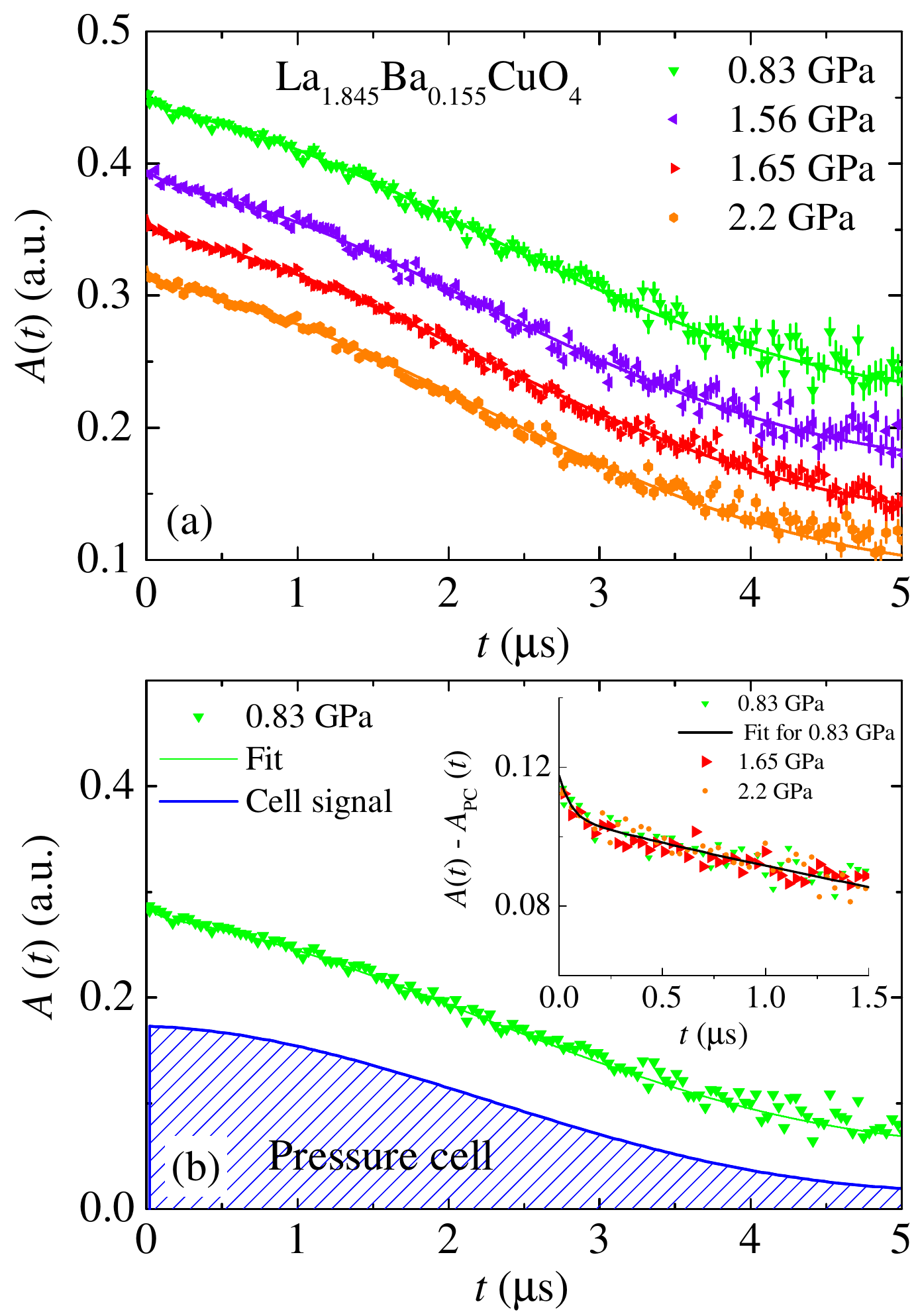}
\vspace{-0.55cm}
\caption{ (Color online) Extracting the sample response from the total ${\mu}$SR signal for LBCO-0.155. (a) ZF ${\mu}$SR time spectra $A(t)$ for LBCO-0.155 measured in the pressure range 0.83 GPa ${\leq}$ $p$ ${\leq}$ 2.2 GPa recorded at $T$ = 5 K. The solid lines represent fits to the data by means of Eqs.~(1) and (2) of the main text. (b) For comparison 
the total signal $A_{tot}(t)$ (green triangles) together with the simulated pressure cell signal $A_{PC}(t)$ (blue solid curve)
is plotted for $p$ = 0.83 GPa. The inset shows the difference between the total and the pressure cell signals [$A_{S}(t)$ = $A(t)$-$A_{PC}(t)$)] for part of the data ($t$ ${\leq}$ 1.5 ${\mu}$s) displayed in panel (b).}
\label{Supfig2}
\end{figure}

 As already mentioned in the main text, in the pressure range 0.83 GPa ${\leq}$ $p$ ${\leq}$ 2.2 GPa, 
instead of the oscillatory behavior seen in the spin-ordered state for $p$ ${\textless}$ 0.67 GPa, a rapidly depolarizing and a weak relaxing ZF-${\mu}$SR signal are observed (see Fig.~\ref{Supfig2}a). The signal with weak
exponential depolarization is affected by a substantial contribution arising from the pressure cell (see Fig.~\ref{Supfig2}b), which is subtracted to extract the sample signal. The resulting ${\mu}$SR signal of the
sample at $p$ = 0.83 GPa is shown in the inset of Fig.~\ref{Supfig2}b.

\end{document}